\documentclass[submit]{epsv8}
\usepackage{graphicx}

\usepackage{amsmath}
\usepackage{amssymb}
\usepackage{bm}
\usepackage{url}
\newcommand{\argmin}{\mathop{\rm arg~min}\limits}

\title{HypoNet Nankai PS: rapid hypocenter determination for the Nankai Trough using physics-informed neural networks with three-dimensional P- and S-wave velocity models}
\author{
Ryoichiro Agata, Japan Agency for Marine-Earth Science and Technology (JAMSTEC), Japan,
agatar@jamstec.go.jp\\
Ayako Nakanishi, Japan Agency for Marine-Earth Science and Technology (JAMSTEC), Japan,
ann@jamstec.go.jp\\
Dan Bassett, Earth Sciences New Zealand, New Zealand,
dan.bassett@earthsciences.nz\\
Satoru Baba, Faculty of Sciences, Kyushu University, Japan,
baba.satoru.083@m.kyushu-u.ac.jp\\
Yasuyuki Nakamura, Japan Agency for Marine-Earth Science and Technology (JAMSTEC), Japan,
yasu@jamstec.go.jp\\
Gou Fujie, Japan Agency for Marine-Earth Science and Technology (JAMSTEC), Japan,
fujie@jamstec.go.jp
}

\abstract{
Rapid and accurate hypocenter determination based on three-dimensional (3D) seismic velocity models is essential for seismic monitoring in subduction zones, yet it remains computationally demanding. HypoNet Nankai, a hypocenter determination package that uses physics-informed neural network (PINN) travel-time inference for a realistic 3D P-wave velocity model of the Nankai Trough subduction zone, was designed to address this problem, but supported only P-wave arrival times. Here, we present HypoNet Nankai PS, which extends the framework to both P- and S-wave travel times. We first constructed an S-wave velocity model from the existing P-wave model using empirical conversion relations for marine sediments and the crust. Because the resulting S-wave velocity structure has lower velocities and therefore considerably steeper travel-time gradients, the original PINN training procedure designed for the P-wave velocity structure was insufficient. We therefore revised the training procedure by reconsidering the distribution of collocation points, the NN architecture, and the optimizer. Benchmark comparisons with the fast marching method showed that the revised approach substantially reduced the root-mean-square deviations for both P- and S-waves compared with the original algorithm. Recovery tests using the DONET and N-net station geometries for synthetic earthquakes showed that the travel-time predictions of HypoNet Nankai PS were sufficiently accurate to reliably recover hypocenters and that the system performed robustly. 
When applied to 223 earthquakes recorded by DONET stations off the Kii Peninsula, the system yielded hypocenters generally consistent with the JMA catalog beneath the network, suggesting that the 3D velocity model captures travel-time variations accommodated by JMA's 1D model and station corrections.
Farther seaward, the estimated hypocenters were substantially shallower than those in the JMA catalog and were located in the oceanic crust and uppermost mantle, consistent with previous relocations that used DONET data and 3D velocity models.
Together, these results establish a compact and flexible framework for rapid hypocenter determination using P- and S-wave travel times in realistic 3D subduction-zone structures, enabling newly deployed or temporary stations to be incorporated without recomputing travel-time tables.
}

\keywords{hypocenter determination; physics-informed neural network; Nankai Trough; travel time; P wave; S wave; DONET}

\begin{document}

\maketitle

\section{Introduction}
\label{sec:introduction}

Accurate hypocenter determination from seismic arrival-time data remains challenging
in seismically active regions such as the Nankai Trough, Southern California, and the Hikurangi margin,
despite its importance for both seismic hazard mitigation and scientific research.
Incorporating realistic three-dimensional (3D) velocity models can substantially improve
hypocenter location accuracy, particularly in structurally complex regions such as subduction
zones \citep{Johnson2002BSSA,Katsumata2015BSSA,Nakano2015}.
Although constructing an appropriate 3D velocity model is demanding, several models
suitable for such analyses are available (e.g., \citealp{Nakanishi2018,Arnulf2022NatGeo,Bassett2022JGR-SE}
for the Nankai Trough, \citealp{Shaw2015EPSL} for Southern California, and
\citealp{Bassett2025JGR-SE} for Hikurangi).
A variety of numerical approaches for 3D seismic travel-time computation have been developed,
including methods based on the eikonal equation \citep{Sethian1996,White2020SRL,Giroux2021}
and ray tracing \citep{Julian1977JG,Um1987BSSA,Moser1991Geophysics}. These techniques have
also been incorporated into software frameworks for determining and relocating hypocenters
in 3D heterogeneous velocity structures \citep{Lomax2000,Zhang2003}.
Nevertheless, hypocenter determination based on one-dimensional (1D) velocity models remains common in practice because constructing 3D velocity models and repeatedly computing travel times through them remain costly, which can hinder the use of 3D models in operational earthquake monitoring and rapid response.

In this context, approaches based on physics-informed neural networks (PINNs)
\citep{Raissi2019} have recently emerged as a promising alternative to traditional
travel-time computation methods.
In a PINN-based framework, a neural network (NN) takes the coordinates of a source and
receiver as inputs and is trained to approximate the solution of a governing equation by
minimizing residuals evaluated at sampled collocation points
\citep{Smith2021,Waheed2021PINNeik}.
Training over all possible source and receiver positions within the target velocity model enables an NN to rapidly infer travel times for arbitrary source--receiver pairs.
PINNs have increasingly been applied to realistic velocity structures and integrated
into hypocenter determination systems \citep{Smith2022,Taufik2023SciRep,Ross2026JGR-SE}.
HypoNet Nankai \citep{Agata2026SRL} is one such tool and is publicly available as a hypocenter determination package that incorporates PINN-based travel-time inference using a realistic 3D
P-wave velocity model of the Nankai Trough region.
HypoNet Nankai is the first PINN-based travel-time modeling and hypocenter determination
system developed for an entire subduction zone that can generate M8-class earthquakes.
However, HypoNet Nankai did not support S-wave velocity structures, primarily because the
S-wave velocity model corresponding to the P-wave velocity model used for PINN training was not available.

Although recent studies have estimated offshore S-wave velocity structures in limited parts of the Nankai Trough \citep{Tonegawa2017NatComms,Liu2025PNAS}, an observationally constrained 3D S-wave velocity model covering the entire region has yet to be constructed. 
Deriving an S-wave velocity model from an existing P-wave model using empirical conversion relations therefore remains a practical approach for regional-scale applications. 
Unconsolidated marine sediments in the shallowest parts of sediment-rich subduction margins such as the Nankai Trough are known to have high Poisson's ratios, often exceeding 0.4, and correspondingly high $V_{\mathrm{P}}/V_{\mathrm{S}}$ ratios that can reach several times the Poisson-solid value of $\sqrt{3}$ \citep{Hamilton1979JASA,Takahashi2002JGR-SE,Liu2025PNAS}.
The resulting low S-wave velocities pose a particular challenge for PINN-based travel-time modeling because, according to the eikonal equation, the magnitude of the spatial travel-time gradient equals the slowness.
The low velocities in the shallow parts of the S-wave velocity model therefore yield substantially steeper travel-time gradients than those in the corresponding P-wave model, making the S-wave travel-time field more difficult for an NN to approximate accurately.
Consequently, a training strategy that accurately represents the P-wave travel-time field does not necessarily provide comparable accuracy for the S-wave field.
Extending a P-wave-based framework to S waves therefore requires reevaluating the training strategy rather than simply training an additional network with the same configuration.

Addressing these challenges, we developed HypoNet Nankai PS, which supports both P- and S-wave travel-time inference.
We first derived an S-wave velocity model from the existing P-wave velocity model using empirical conversion relations that account for the high Poisson's ratios of marine sediments.
We then developed a PINN model capable of inferring travel times for arbitrary source--receiver pairs within this S-wave velocity structure.
To achieve accurate PINN-based travel-time inference for the S-wave velocity structure, which has lower velocities and therefore steeper travel-time gradients than the P-wave structure, we revised the training procedure by sampling both source and collocation points from the full 3D domain with a random-distance sampling strategy, adopting the Deeper-PINN architecture, and replacing the Adam optimizer with the second-order SOAP optimizer.
We also retrained the PINN for P-wave travel times using the updated training configuration.
We assessed the accuracy of the resulting P- and S-wave travel-time models through benchmark comparisons with the fast marching method (FMM) \citep{Sethian1996}. We then conducted comprehensive synthetic hypocenter determination tests to evaluate whether the remaining differences between the PINN and FMM travel times affect location accuracy, assess the robustness of the determination system, and examine whether the results reflect the inherent limitations of the station geometry. Finally, we demonstrated its practical applicability by determining hypocenters from observed P- and S-wave arrivals recorded by the Dense Oceanfloor Network system for Earthquakes and Tsunamis (DONET) \citep{Kaneda2015,Aoi2020} ocean-bottom seismometers off the Kii Peninsula. 
Ultimately, HypoNet Nankai PS aims to enable a broad range of
researchers to readily obtain reliable hypocenter distributions based
on 3D velocity structures as the first step toward more detailed
analyses, including for datasets from newly deployed and temporary
stations.

\section{Construction of S-wave Velocity Model of the Nankai Trough Subduction Zone}
\label{sec:velocity}

\begin{figure}[htbp]
\centering
\includegraphics[clip, width=\textwidth]{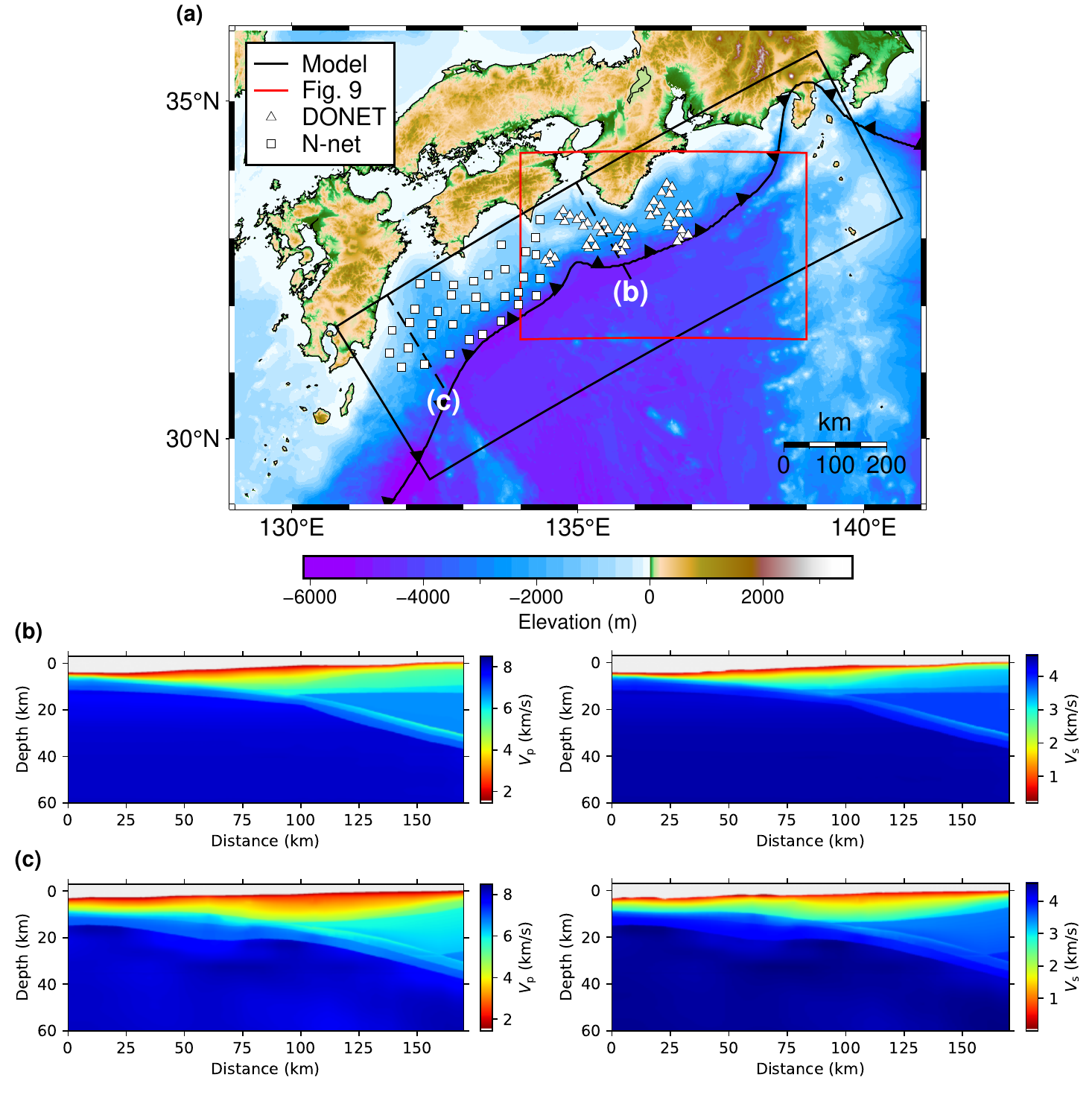}
\caption{(a) Map of the study region showing the PINN modeling domain and target regions for hypocenter determination. Lines labeled (b) and (c) indicate the locations of the velocity cross sections.
(b, c) Cross sections of the N2018 P-wave velocity model (left) and the S-wave velocity model derived by empirical conversion (right) along the profiles marked in (a).}
\label{fig:map}
\end{figure}

To develop an NN model for S-wave travel-time calculations, we first constructed an S-wave velocity model of the Nankai Trough subduction zone based on the P-wave velocity model proposed by \citet{Nakanishi2018} (hereafter referred to as the N2018 model).
The N2018 model was constructed by combining two-dimensional (2D) P-wave velocity profiles derived from previous wide-angle seismic reflection surveys that used OBS arrival-time data \citep{JAMSTEC2004} with a 3D velocity model derived from seismic tomography of natural earthquakes \citep{Yamamoto2013,Yamamoto2014}. Following \citet{Agata2026SRL} (hereafter referred to as A2026), we focused on developing PINNs for travel-time calculations in the offshore region. This region covers a horizontal area of 900\,km$\times$300\,km and extends to a depth of 60\,km (Fig. \ref{fig:map}(a)). We selected this region because the N2018 model prioritizes accurate modeling of offshore structures.

We used the following empirical relationships between P- and S-wave velocities to construct the S-wave velocity model from the N2018 model \citep{MEXT2022NankaiBosai}:
For marine sediments extending from the seafloor to 500 m below the seafloor, we used conversion relations derived from in situ measurements \citep{Hamilton1979JASA}. At greater depths, we used crustal relations derived from a compilation of borehole logs, vertical seismic profiles, laboratory measurements, and seismic tomography models \citep{Brocher2005}.
Although Hamilton's relation is not applicable to P-wave velocity values lower than 1.512\,km/s, the N2018 model contains values slightly below this limit (but still above the typical sound speed in seawater of 1.5\,km/s), resulting in unrealistically low S-wave velocities.
Based on the range of S-wave velocities previously estimated for the Nankai Trough region using seismic data \citep{Tonegawa2017NatComms,Liu2025PNAS}, we set a minimum threshold of 0.2\,km/s for the S-wave velocity model. 
Applying this threshold changed the velocity in less than 0.1\,\% of the total model volume.
As a result, the largest $V_{\mathrm{P}}/V_{\mathrm{S}}$ ratio in our velocity models was around 7.7. 
Figures \ref{fig:map}(b) and (c) show cross sections through the N2018 P-wave and newly developed S-wave velocity models.

We transformed the velocity-model data into a topocentric Cartesian coordinate system centered at (136$^{\circ}$E, 33$^{\circ}$N, 0), without using a map projection.
We used geoid data from EGM2008 \citep{Pavlis2012} for this transformation.
We further translated and rotated the coordinate system to align it with the grid used for subsequent travel-time comparisons.
In the original N2018 dataset, point-cloud locations had uniform spacing only in the vertical direction.
On average, the N2018 model sampled points at 0.1\,km intervals vertically and approximately 0.5\,km intervals horizontally.
Although dense vertical sampling helps capture sharp velocity discontinuities, such as the Moho, it oversamples other regions with moderate velocity changes.
In regions where the vertical velocity gradient was less than $0.1\,\mathrm{s}^{-1}$, we downsampled the point cloud to a grid spacing of 1\,km.
This approach reduced computational cost by concentrating resolution in regions with significant vertical velocity changes.
After this procedure, the dataset contained approximately 50 million points. We used all of these points for mini-batch training of the PINN, as described later.
Because grid-based travel-time calculation methods provide reference solutions for validating our calculations, we did not split the data into validation and test sets, unlike standard machine-learning practice.

We used the GEBCO Gridded Bathymetry Data \citep{Gebco2023} and EGM2008 geoid data \citep{Pavlis2012} to define the Earth's surface within the velocity-model domain.
As with the velocity model, we transformed the original bathymetric data, which consisted of latitude, longitude, and elevation, into the Cartesian coordinate system.

\section{PINN-based travel-time calculation and hypocenter determination}
\label{sec:PINN_training_hypocenter}

\subsection{Theory of PINN for travel-time prediction}
\label{sec:PINN_theory}

We describe the PINN formulation for travel-time calculation \citep{Smith2021,Waheed2021PINNeik,Grubas2023}, which is a slight modification of the formulation proposed by \citet{Agata2023}.
The eikonal equation relates the spatial derivative of the travel-time field to the velocity structure as follows:
\begin{align}
|\nabla T(\mathbf{x},\mathbf{x}_\mathrm{s})|^{2} &= \displaystyle \frac{1}{v^{2}(\mathbf{x})}, \quad \forall\,\mathbf{x} \in \Omega \label{eqn:eikonal}\\
T(\mathbf{x}_\mathrm{s}, \mathbf{x}_\mathrm{s}) &= 0,\label{eqn:PC}
\end{align}
where $\Omega$ is a domain in $\mathbb{R}^d$, $d$ is the spatial dimension, $T(\mathbf{x},\mathbf{x}_\mathrm{s})$ is the travel time from the source $\mathbf{x}_s$ to the point $\mathbf{x}$, $v(\mathbf{x})$ is the velocity at $\mathbf{x}$, and $\nabla$ denotes the gradient operator with respect to $\mathbf{x}$.
The second equation defines the point-source condition.
We impose the source condition as a hard constraint through multiplicative factorization \citep{Waheed2021PINNeik}.
We also apply additional operations to normalize the inputs and outputs of the NNs, improve convergence, and constrain the final output values within prescribed upper and lower limits.
Furthermore, we impose the reciprocity condition (i.e., $T(\mathbf{x},\mathbf{x}_\mathrm{s})=T(\mathbf{x}_\mathrm{s},\mathbf{x})$) following \citet{Grubas2023} to improve convergence to the solution of the eikonal equation.

\begin{figure}[htbp]
\centering
\includegraphics[clip, width=\textwidth]{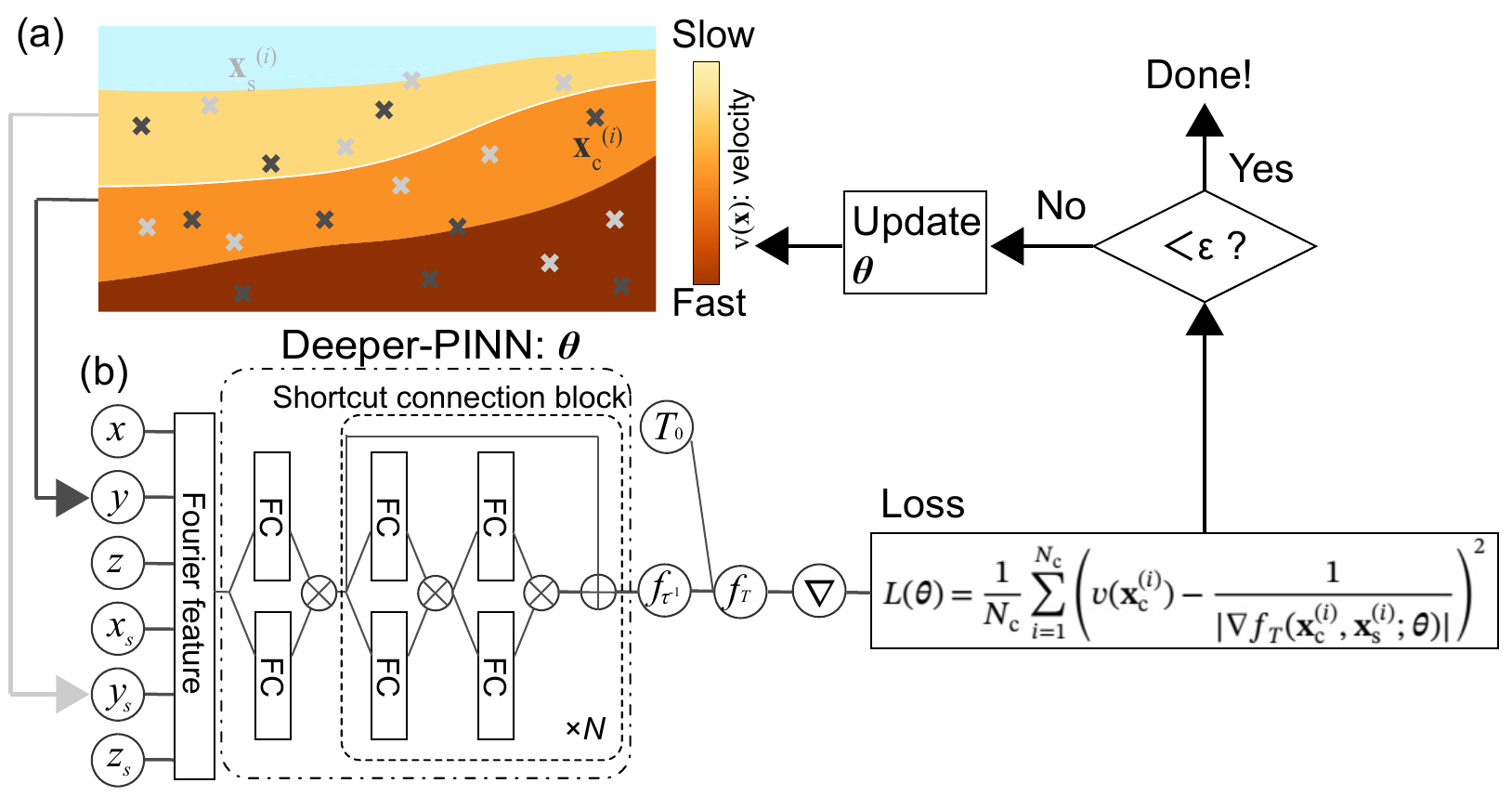}
\caption{(a) Schematic of PINN-based travel-time learning.
(b) Deeper-PINN architecture used in this study, which employs a random Fourier feature embedding.}
\label{fig:NN}
\end{figure}

We train the NNs using the following loss function:
\begin{equation}
L(\boldsymbol{\theta}) = \frac{1}{N_{\rm c}} \sum_{i=1}^{N_{\rm c}}\left(v(\mathbf{x}_{\rm c}^{(i)})-\frac{1}{|\nabla f_{T}(\mathbf{x}_{\rm c}^{(i)},\mathbf{x}_{\rm s}^{(i)}; \boldsymbol{\theta})|} \right)^{2}, \label{eqn:loss}
\end{equation}
where $N_c$ is the number of collocation-point/source-point pairs.
$\mathbf{x}_{\rm c}$ and $\mathbf{x}_{\rm s}$ are the coordinates of the collocation and source points, respectively. 
The collocation points are the points at which the PDE residuals are evaluated \citep{Raissi2019}.
The loss function incorporates physics-informed constraints and consists of the sum of the squared residuals for each source-point/collocation-point pair.
The travel-time gradient needed to evaluate the residual can be readily obtained using automatic differentiation (Figure \ref{fig:NN}).
The coordinates $\mathbf{x}_{\rm c}^{(i)}$, which correspond to $\mathbf{x}$ in Equation \ref{eqn:eikonal}, must cover the entire domain to ensure that the eikonal equation is satisfied everywhere, regardless of how limited the receiver distribution is.
In contrast, the distribution of $\mathbf{x}_{\rm s}^{(i)}$ in Equation \ref{eqn:loss}, which corresponds to $\mathbf{x}_s$ in Equation \ref{eqn:eikonal}, can be chosen arbitrarily within the domain of interest for sources. 
$f_{T}(\mathbf{x}_{\rm c},\mathbf{x}_{\rm s}; \boldsymbol{\theta}^{*})$ with the optimized NN weight parameters $\boldsymbol{\theta}^{*}= \argmin_{\boldsymbol{\theta}} L(\boldsymbol{\theta})$ can rapidly infer travel times for any in-distribution source--receiver pair.

\subsection{Algorithm and settings of PINN training for travel-time inference}
\label{sec:PINN_training}

Because the minimum velocity in our S-wave model is as low as 0.2\,km/s, the spatial gradients of the S-wave travel-time field are substantially larger than those of the P-wave field, making accurate approximation of the field by the NN more difficult.
Preliminary training experiments confirmed that the algorithm adopted in the original HypoNet Nankai \citep{Agata2026SRL} was insufficient to infer the S-wave travel-time field with acceptable accuracy, necessitating a systematic revision of the training procedure. The updated algorithm consists of three main modifications described below: the sampling strategy for source and collocation points, the optimizer, and the NN architecture.
Table \ref{tab:algorithm} summarizes the differences in training settings between A2026 and HypoNet Nankai PS.
We also applied the updated algorithm to retrain the P-wave PINN.

We sampled both $\mathbf{x}_{\rm c}^{(i)}$ and $\mathbf{x}_{\rm s}^{(i)}$ from the full 3-D domain to improve inference accuracy.
This differs from A2026, in which the source points $\mathbf{x}_{\rm s}^{(i)}$ were drawn exclusively from the Earth's surface, whereas the collocation points $\mathbf{x}_{\rm c}^{(i)}$ were sampled from the full 3-D velocity-model domain.
In A2026, this strategy interchanged the source and receiver roles at inference time using reciprocity, allowing the NN to predict travel times from any subsurface source to any surface receiver.
Although restricting $\mathbf{x}_{\rm s}^{(i)}$ to the Earth's surface, i.e., a 2D manifold, narrows the solution space and thereby appears to facilitate training, our experiments presented later demonstrate that sampling both source and receiver points throughout the full 3-D domain combined with the random-distance sampling  strategy \citep{Smith2021} yields higher travel-time inference accuracy: 
The random-distance sampling provides a more uniform distribution of source--receiver distances by first sampling a source location and then placing the receiver at a random distance along a random direction.
A new set of collocation points was randomly sampled in each epoch and used for a single loss evaluation and parameter update.


Learning a travel-time function over the six-dimensional input space formed by the coordinates of a source and a receiver, each specified in 3-D, requires a network with sufficient capacity.
We adopted the Deeper-PINN architecture \citep{Jiang2025ASC}, which was specifically designed so that inference accuracy improves reliably as network depth increases, even when gradients of the NN output appear in the loss function, as in a PINN-based loss (e.g., Equation \ref{eqn:loss}).
Deeper-PINN extends a standard fully connected PINN by stacking multiple shortcut connection blocks to improve the trainability and expressiveness of deep networks. In this study, each fully connected layer had a width of 256, and the network consisted of 30 shortcut connection blocks.
We retained the random Fourier feature (RFF) embedding \citep{Tancik2020,Hennigh2021} used in A2026 as the input encoding layer.
Domain decomposition, which was employed in A2026 to achieve accurate travel-time inference within local subdomains, was not used in the present study. 
A single Deeper-PINN handling the entire domain yielded better inference accuracy, as shown in Section \ref{sec:forward}.
Using a single domain rather than domain decomposition affects only the accuracy and does not change the user experience.

To accelerate training of this larger network, we adopted the state-of-the-art second-order optimizer SOAP (Shampoo as Adam's Preconditioner) \citep{Vyas2025ICLR}, which has been reported to improve the convergence of PINNs \citep{Wang2025NeurIPS}.
SOAP combines hidden layer-wise second-order preconditioning with an efficient Adam-style update rule, providing faster convergence than Adam \citep{Kingma2015} for NN training.
We selected the P- and S-wave models at 30,000 and 40,000 epochs, respectively, trained by using eight NVIDIA A100 40GB GPUs in Earth Simulator at JAMSTEC.
We terminated the training based on the root-mean-square deviation (RMSD) of inferred travel times with FMM results (see Section \ref{sec:forward}). 



\subsection{Hypocenter determination method}
\label{sec:hypocenter}

HypoNet Nankai PS uses a maximum a posteriori (MAP) approach for hypocenter determination \citep{Hirata1987}, following A2026.
This method defines a joint posterior probability density function (PDF) for the hypocenter parameters (longitude, latitude, and depth) and the origin-time shift. 
The joint PDF is marginalized over the origin time, yielding the posterior PDF for the hypocenter locations alone.
The MAP solution of this location posterior is obtained using the gradient-based L-BFGS method \citep{Liu1989}. To reduce sensitivity to the initial hypocenter location, the system performs three parallel searches starting beneath the station with the earliest observed arrival, at depths of 7.7, 30.5, and 53.3\,km, and selects the solution with the highest posterior probability.
Location uncertainties are estimated using the Laplace approximation around the MAP solution.
The origin-time shift and its uncertainty are additionally estimated as a post-processing step.

Whereas the hypocenter determination system presented in A2026 was designed primarily for CPU-based operation, we implemented the present system to take advantage of GPU acceleration. Although the achievable speedup depends on the number of observations and the hardware specifications, GPU execution is expected to reduce the computation time for hypocenter determination by a factor of several.
The prior distribution, optimization, parameterization, and other settings generally follow A2026. 

\section{Numerical validation experiments}
\label{sec:verification}

\subsection{Forward travel-time comparison with FMM}
\label{sec:forward}

We designed verification problems to compare travel times inferred by the trained PINN with those computed using FMM \citep{Sethian1996,Ganster2023eikonalfm}, one of the most widely used grid-based finite-difference methods.
We generated 20 random seismic sources within the target 3D domain and computed travel times at virtual receiver points spaced at 1\,km intervals on the Earth's surface.
The FMM grid intervals were 0.5\,km horizontally and 0.1\,km vertically, matching those of the data points in the N2018 velocity model.
Preliminary runs confirmed that these grid intervals were sufficiently fine to provide reliable reference solutions.
We used 15 sources as a validation set to determine the training termination epochs reported in Section \ref{sec:PINN_training}. Every 10,000 epochs, we calculated the root-mean-square difference (RMSD) between the PINN and FMM travel times over all surface grid points for each validation source and used the mean of these per-source RMSDs as the validation metric (Figure S1). The remaining five sources were reserved as a test set.

\begin{figure}[htbp]
\centering
\includegraphics[clip, width=\textwidth]{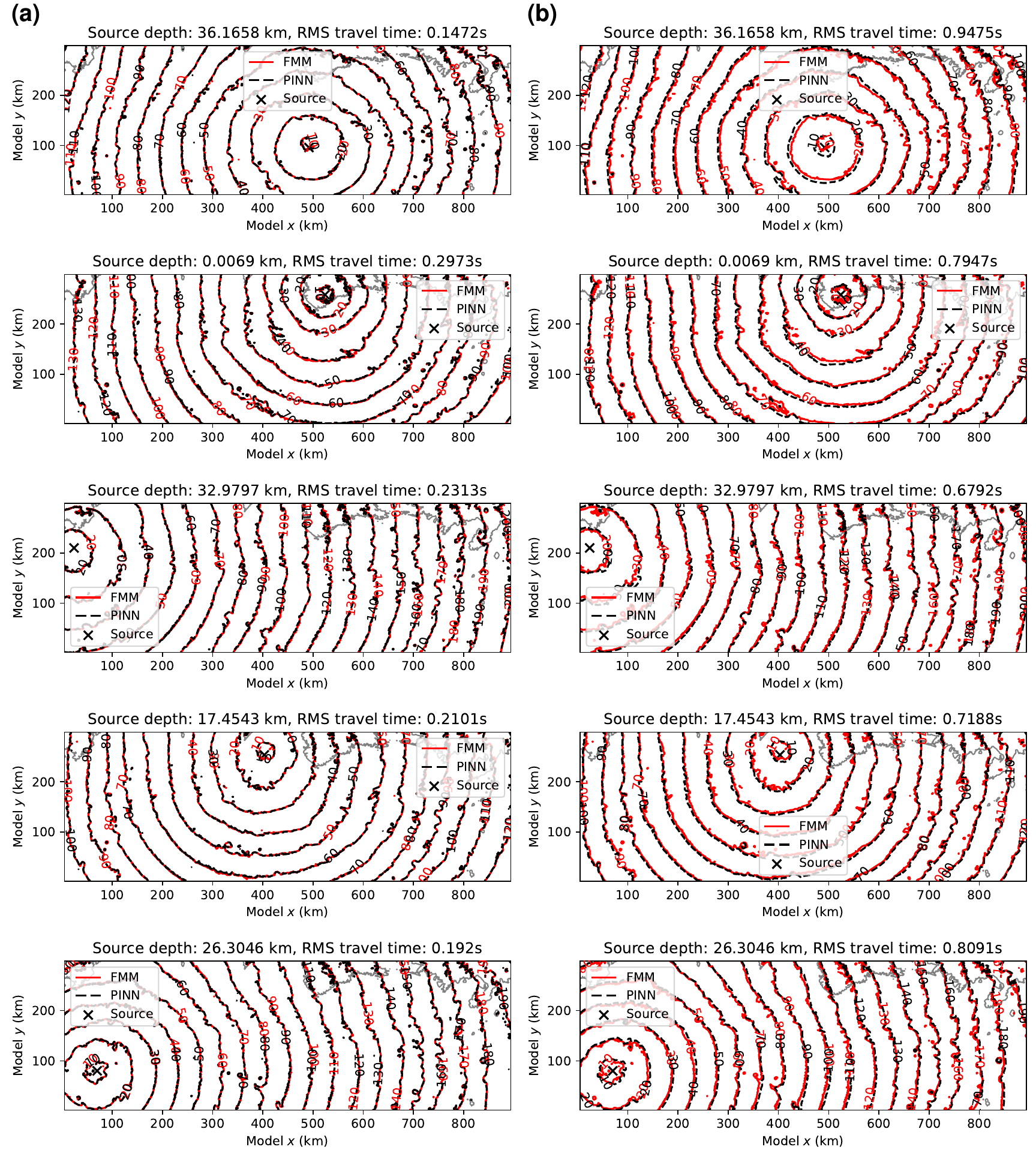}
\caption{Comparison of S-wave travel-time fields on the Earth's surface inferred by the PINN and computed by FMM for five randomly selected sources.
(a) Results obtained with the training algorithm used in this study.
(b) Results obtained with the training algorithm of A2026 without domain decomposition.
The epicenter location, source depth, and RMSD between the PINN and FMM results are indicated in each panel.}
\label{fig:forward_s}
\end{figure}

\begin{figure}[htbp]
\centering
\includegraphics[clip, width=0.5\textwidth]{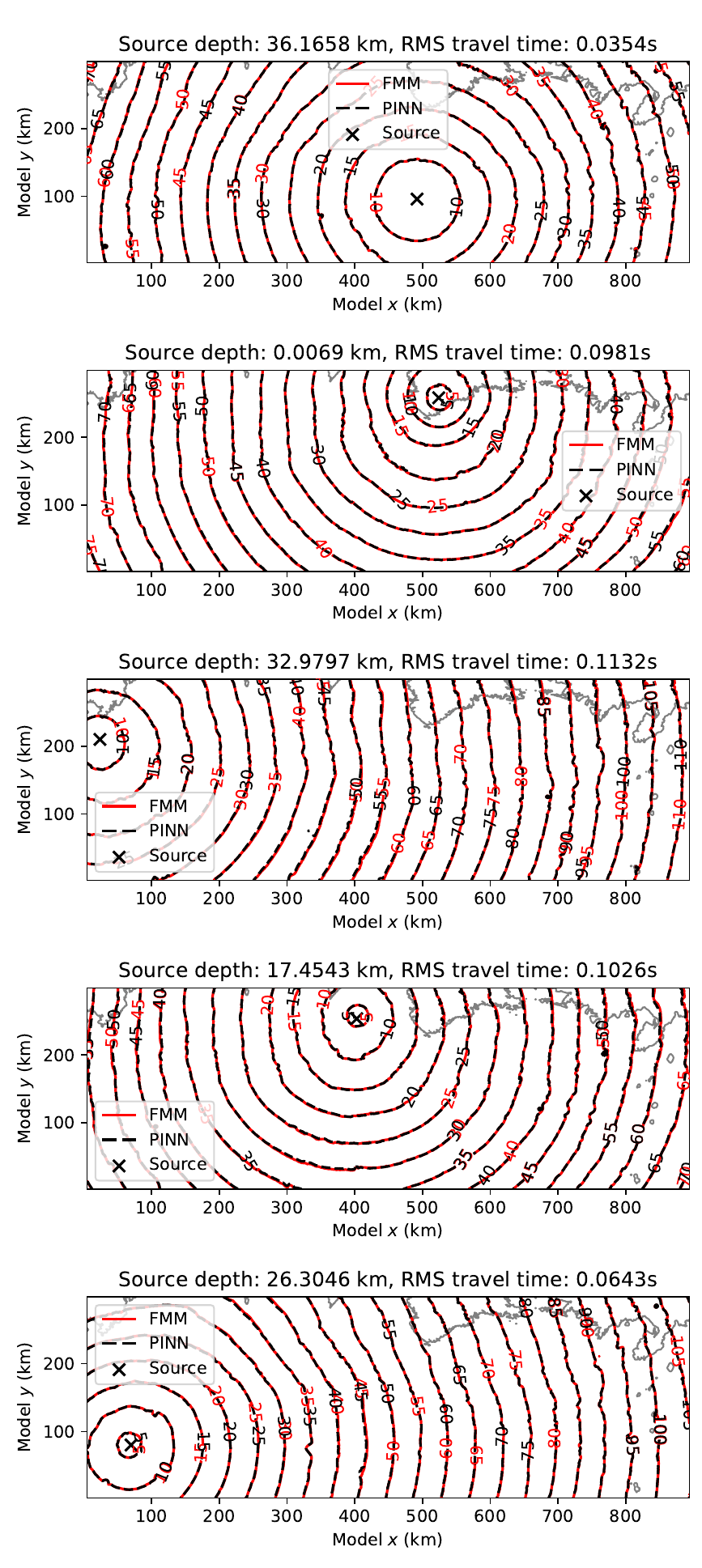}
\caption{Comparison of P-wave travel-time fields on the Earth's surface inferred by the PINN trained with the algorithm used in this study and computed by FMM for five randomly selected sources.
The epicenter location, source depth, and RMSD between the PINN and FMM results are indicated in each panel.}
\label{fig:forward_p}
\end{figure}

For the S waves, the travel-time functions on the Earth's surface for five test sources, calculated using both PINN and FMM, show good agreement (Fig. \ref{fig:forward_s}(a)).
We quantified the difference using RMSD at the surface grid points.
The RMSDs for S-wave travel-time inference range from 0.1 to 0.3\,s.
We also computed travel times using an NN trained with the same algorithm as in A2026, which resulted in considerably larger RMSDs, ranging from 0.6 to 1.0\,s (Fig. \ref{fig:forward_s}(b)). 
For simplicity, we omitted the domain decomposition used in A2026, which had little impact on the overall evaluation. 
We attribute the improved travel-time accuracy to the newly introduced PINN training strategy presented in Section \ref{sec:PINN_training}.
P-wave travel times from PINN and FMM agree with RMSDs smaller than 0.12\,s, compared with those ranging from 0.1 to 0.3\,s in A2026, suggesting improved accuracy with the present training strategy (Fig. \ref{fig:forward_p}). 

\begin{figure}[htbp]
\centering
\includegraphics[clip, width=\textwidth]{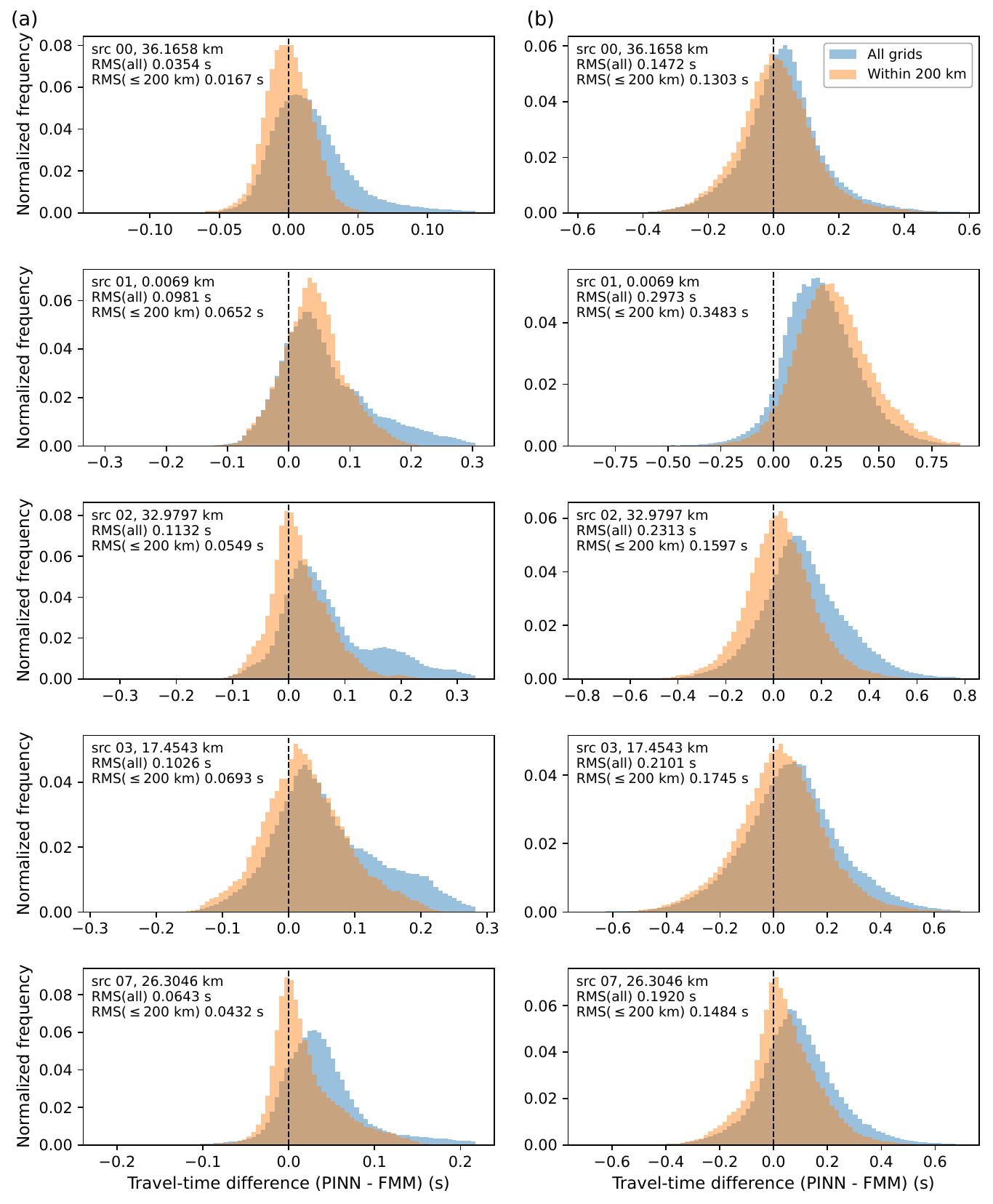}
\caption{Histograms of the travel-time difference (PINN minus FMM) at surface grid points for the five sources shown in Figures \ref{fig:forward_s} and \ref{fig:forward_p}.
(a) P-wave results.
(b) S-wave results.}
\label{fig:histogram}
\end{figure}

We further visualized the distribution of the differences between travel times inferred by the PINN trained in this study and those computed with FMM at each grid point on the Earth's surface using histograms (Fig. \ref{fig:histogram}).
In addition to the results for all grid points, we compared the results for points within 200\,km of the source in the horizontal direction. 
We made this comparison because the hypocenter determinations routinely performed by the Japan Meteorological Agency (JMA) place greater weight on stations closer to the epicenter. 
The histograms show that the RMSD is generally smaller when only surface points within 200\,km of the source are evaluated although their shapes vary among sources.

\subsection{Synthetic hypocenter determination}

The results in the previous subsection validate the ability of our PINN to provide travel times of comparable accuracy to FMM for forward modeling.
However, we further investigated how differences between the two methods affected the target inverse problem of hypocenter determination. We also assessed the robustness of our determination algorithm. In addition, we examined whether the results were consistent with physical expectations regarding the inherent limitations imposed by limited observational coverage. 

We conducted synthetic hypocenter-determination tests using P- and S-wave arrival times, assuming the use of seafloor seismic observation systems in the Nankai Trough subduction zone.
We used 51 stations of DONET \citep{Kaneda2015,Aoi2020} and 36 stations of Nankai Trough Seafloor Observation Network for Earthquakes and Tsunamis (N-net) \citep{Aoi2023IEEE-UT}, which together span the major portion of the offshore model domain (Fig. \ref{fig:map}(a)).
To perform the test comprehensively, we placed synthetic hypocenters on a geographic grid covering this domain, with a horizontal spacing of approximately 20\,km and at five ellipsoidal heights of -10, -20, -30, -40, and -50\,km.
In total, we considered 3313 sources whose Cartesian coordinates fell inside the computational domain.
We calculated synthetic travel-time data between these hypocenters and the DONET and N-net stations using FMM and used these values as synthetic observations.
We added Gaussian noise with a variance of $(0.1\,\mathrm{s})^{2}$ to the synthetic travel-time data.
We then performed hypocenter determination using HypoNet Nankai PS to estimate the source locations.
Each hypocenter estimate was completed in less than 1\,s using an NVIDIA RTX PRO 6000 GPU. This computation time is sufficiently short for real-time applications, given the specifications of a widely used automatic hypocenter determination system \citep{Matsumura2006Zisin}.

\begin{figure}[htbp]
\centering
\includegraphics[clip, width=\textwidth]{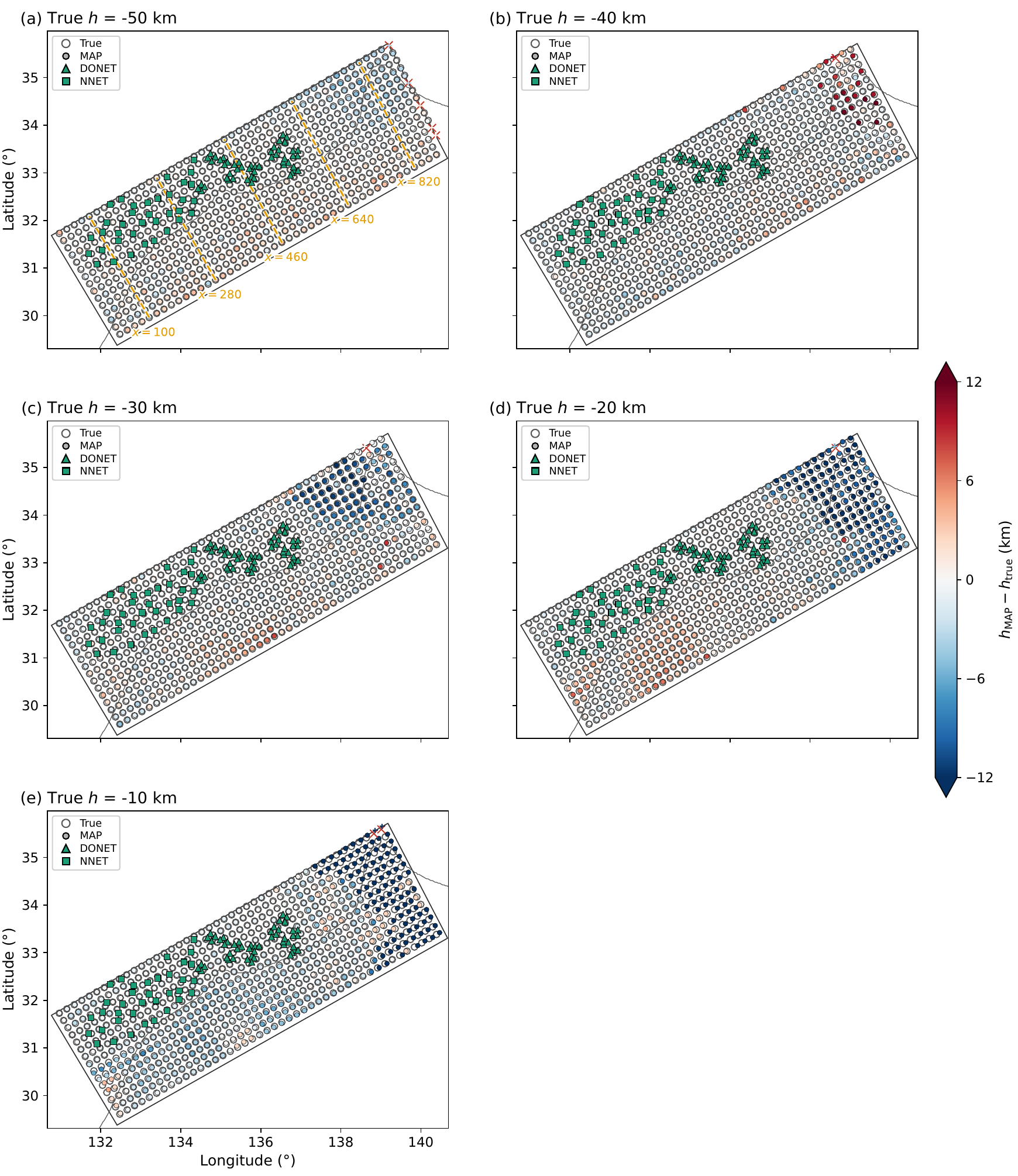}
\caption{Results of synthetic hypocenter determination using P- and S-wave arrival times at virtual DONET and N-net stations.
Map views of MAP-estimated hypocenters are compared with the true locations for true ellipsoidal heights of (a) $-50$\,km, (b) $-40$\,km, (c) $-30$\,km, (d) $-20$\,km, and (e) $-10$\,km.
Colors indicate the residual in ellipsoidal height ($h_{\mathrm{MAP}}-h_{\mathrm{true}}$). 
Red crosses mark the true epicenters of events whose recovered locations fell outside the modeling domain.
In (a), the locations of cross sections presented in Fig. \ref{fig:hypo_yh} are indicated.}
\label{fig:hypo_lonlat}
\end{figure}

\begin{figure}[htbp]
\centering
\includegraphics[clip, width=\textwidth]{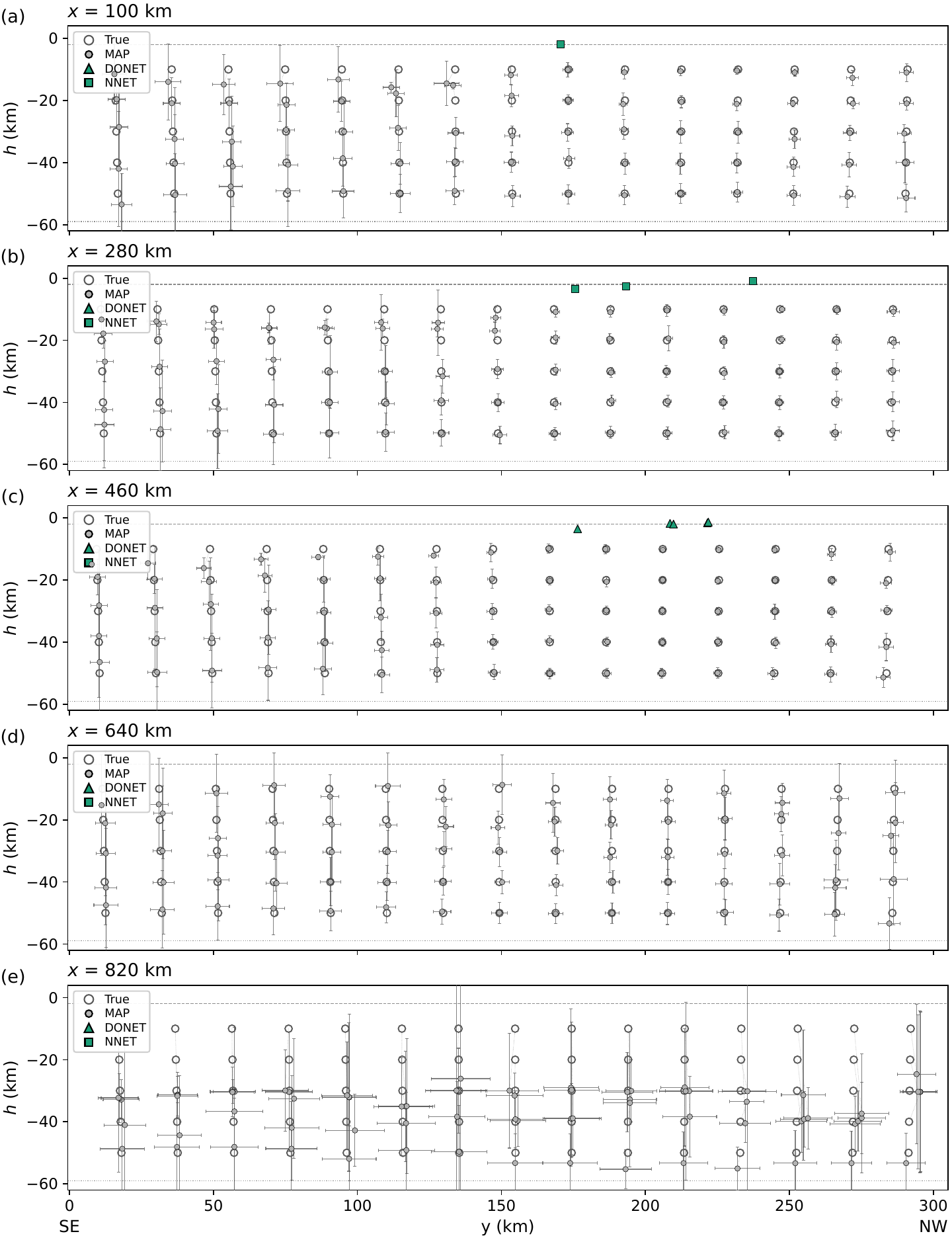}
\caption{Cross sections of the synthetic hypocenter determination results shown in Fig. \ref{fig:hypo_lonlat}, presented in the $y$--$h$ plane at (a) $x=100$\,km, (b) $x=280$\,km, (c) $x=460$\,km, (d) $x=640$\,km, and (e) $x=820$\,km.
True and MAP-estimated locations are compared. Crosses indicate $2\sigma$ confidence intervals for the estimated hypocenter locations. Covariance components are omitted for clarity.
NW and SE below the $y$--axis in (e) denote northwest and southeast, respectively.}
\label{fig:hypo_yh}
\end{figure}

\begin{figure}[htbp]
\centering
\includegraphics[clip, width=12.5cm]{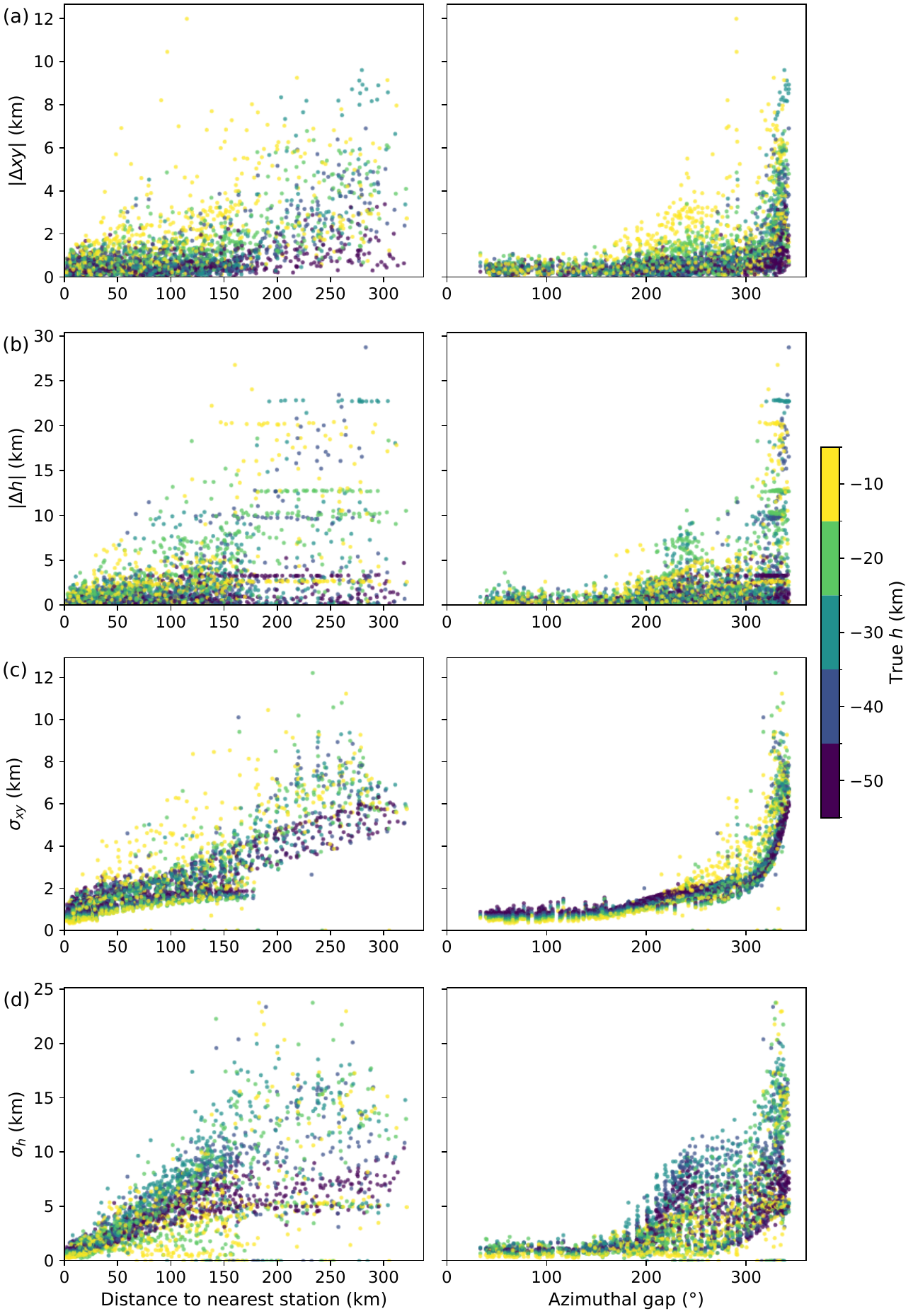}
\caption{Location errors and posterior uncertainties in the synthetic test plotted against the distance to the nearest station (left column) and the azimuthal gap (right column).
The distance is the epicentral (horizontal) distance from the true hypocenter to the nearest DONET or N-net station.
The azimuthal gap is the largest angle between the azimuths from the true epicenter to two adjacent stations.
(a) Absolute horizontal error $|\Delta xy|$ between the MAP and true hypocenters.
(b) Absolute error in ellipsoidal height $|\Delta h|$.
(c) Horizontal $1\sigma$ uncertainty $\sigma_{xy}$ obtained from the Laplace approximation of the posterior PDF (in the model Cartesian frame).
(d) Corresponding $1\sigma$ uncertainty in ellipsoidal height $\sigma_h$.
Symbols are colored by the true ellipsoidal height $h$.}
\label{fig:error_vs_station}
\end{figure}

The MAP-estimated horizontal locations of the hypocenters within and around the DONET and N-net observation networks are generally accurate and precise (Fig. \ref{fig:hypo_lonlat}).
Hypocenters on the eastern side of the modeling domain show moderate horizontal location errors, defined as the distance between the estimated and true hypocenters, reaching approximately 10\,km at most.
For these hypocenters, depth errors are larger, reaching nearly 30\,km at some points.
Cross sections showing the estimation uncertainty illustrate these tendencies more clearly. The sections at $x$=100\,km, 280\,km, and 460\,km (Fig. \ref{fig:hypo_yh} (a)(b)(c)) suggest that the estimated locations of events at 150\,km$\le y \le$300\,km, which are located below the observation network, agree well with the true locations and have small uncertainties, with standard deviations generally below 1.5\,km.
Outside these regions, both location errors and estimation uncertainties tend to be larger.
At the sections at $x$=640\,km and 820\,km (Fig. \ref{fig:hypo_yh}(d)(e)), above which no observation points are located, both location errors and estimation uncertainties are larger throughout.
The extent appears to depend on the distance from the observation points and on their azimuthal coverage.
To clarify these relationships, we further visualized the location errors and estimation uncertainties against the distance from the nearest station and the azimuthal gap (Fig. \ref{fig:error_vs_station}).
The azimuthal gap is defined as the largest angle between the azimuths from the epicenter to two adjacent stations, with larger values indicating poorer station coverage.
Although the location errors exhibit substantial scatter and the effect of the finite extent of the model domain is apparent, the upper bounds of the errors tend to increase with both the station distance and the azimuthal gap.
In contrast, the estimated horizontal and vertical uncertainties show a clear systematic increase with both quantities.
The vertical uncertainty starts to increase at an azimuthal gap of about 200$^\circ$, whereas the horizontal uncertainty remains relatively stable up to about 300$^\circ$ and then increases sharply.

In summary, when information from data is sufficient, the inverted locations are consistent with the true locations despite the difference in the forward solvers used in data generation and inversion, that is, FMM and PINN, respectively.
The dependence of the location errors and estimation uncertainties on the station distance and azimuthal coverage is consistent with the expected effects of station geometry, providing further support that the hypocenter determination system functions appropriately.

\section{Application to real data: Hypocenter relocation using DONET observations off the Kii Peninsula}
\label{sec:application}

The offshore Nankai Trough off the Kii Peninsula, including the Kumano-nada and Tonankai segments, is characterized by the subduction of the Philippine Sea Plate beneath the Eurasian Plate and a highly heterogeneous velocity structure associated with thick accretionary-prism sediments and the subducting oceanic crust.
In such a setting, accounting for three-dimensional velocity heterogeneity is particularly important for accurate hypocenter determination.
To assess the applicability of HypoNet Nankai PS to observed arrival-time data, we applied the system to earthquakes recorded by DONET off the Kii Peninsula. 
Rather than comparing the resulting hypocenters with those reported in previous studies that used different 3D velocity models, we compared them with JMA catalog hypocenters. This comparison is intended for illustrative purposes, with the JMA hypocenters serving solely as an operational reference for interpreting the results. Because the two analyses used different station sets, velocity models, and station-correction schemes, the comparison is not intended to directly assess relative location accuracy or isolate the effect of the 3D velocity model.

\begin{figure}[htbp]
\centering
\includegraphics[clip, width=14cm]{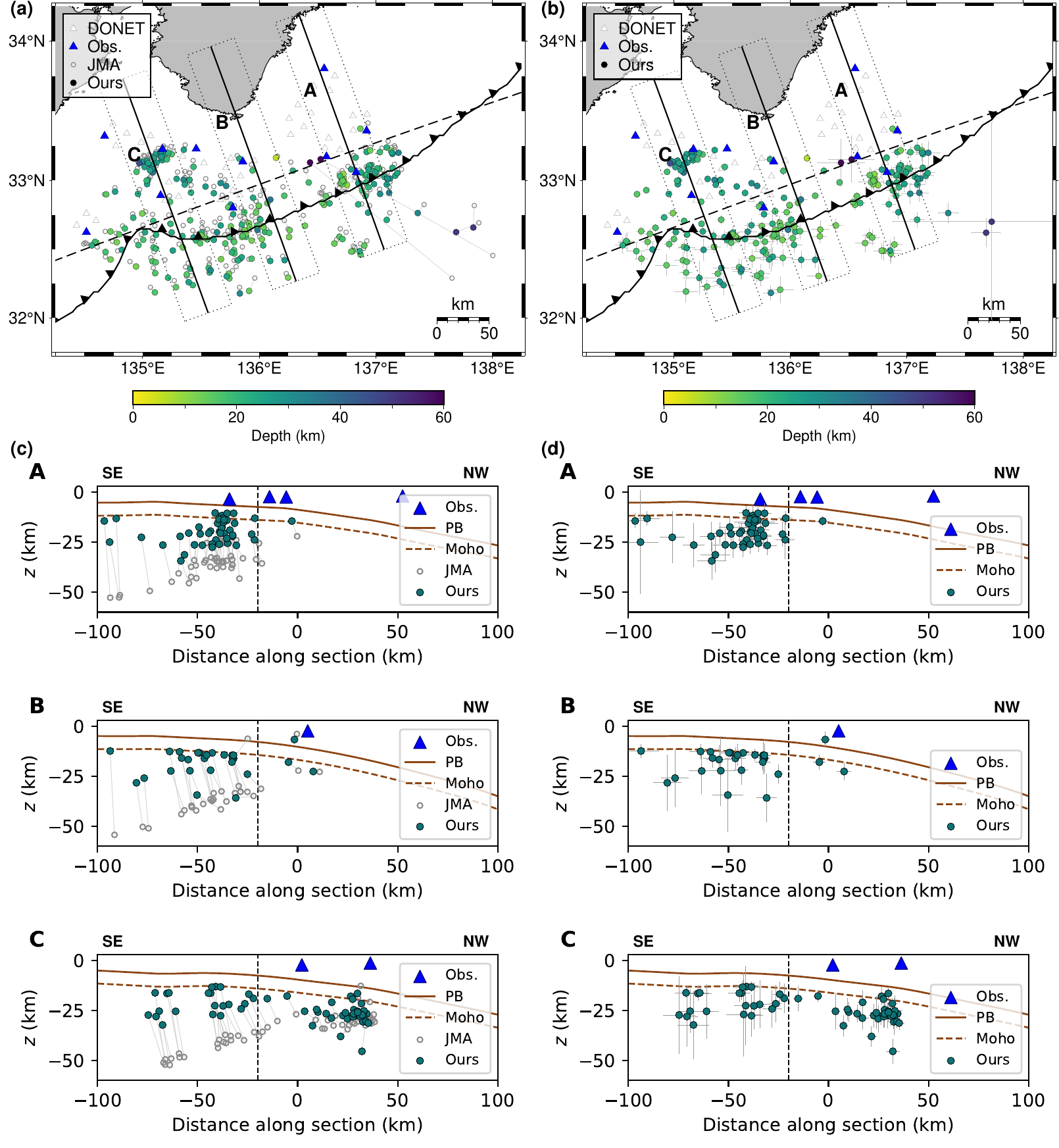}
\caption{Hypocenter relocation using P- and S-wave arrival times at DONET stations.
(a) Map view of MAP-estimated hypocenters compared with the original JMA catalog locations. Colors indicate hypocenter depth. Lines A--C mark the locations of the cross sections.
(b) Same map view as (a), but showing estimation uncertainties. 
(c) Cross sections along A--C showing the changes in hypocenter locations from the JMA catalog.
(d) Same cross sections as (c), but showing estimation uncertainties.
NW and SE in the cross sections denote northwest and southeast, respectively.
The plate boundary (PB) of \citet{Nakanishi2018} and a reference horizon for the Moho, placed 6.5\,km below PB as a visual guide, are indicated in the cross sections. Crosses in (b) and (d) indicate $2\sigma$ confidence intervals. Covariance components are omitted for clarity.
In all the panels, the dashed line indicates a distance of -20\,km along the cross sections, which separates the landward group (distances larger than -20\,km) from the seaward group (distances smaller than -20\,km) in Figure \ref{fig:hypo_depth_section0}.}
\label{fig:DONET_JMA}
\end{figure}

For hypocenter determination, the JMA uses one-dimensional velocity models assigned according to station location: JMA2001A for land stations and JMA2020C for ocean-bottom stations in the Nankai Trough, with station corrections applied to the latter to account for low-velocity sediments beneath the seafloor \citep{JMA_catalog_userguide}.
The JMA determines hypocenters using P- and S-wave arrivals from land and ocean-bottom stations selected from the nationwide observation network.
Between September 2020 and December 2023, the latest period with available records, stations in the entire DONET systems, namely DONET1 and DONET2, were included in the JMA unified processing.
The JMA unified processing included eleven of the 51 stations connected to the twelve DONET nodes (see blue triangles in Fig. \ref{fig:DONET_JMA}(a)). This selection reflects the close proximity of stations deployed at the same node \citep{Naoi2024EPS}.


We selected 223 natural earthquakes whose JMA hypocenters are located within the modeling domain and for which P- or S-wave arrivals were reported at six or more distinct DONET stations in the catalog, allowing stable hypocenter determination.
We did not use land-station arrivals because the target domain of HypoNet Nankai PS is an offshore region.
Accordingly, differences from the JMA catalog reflect differences between the complete analysis configurations rather than the effect of the velocity model alone.
We obtained travel times by subtracting the JMA origin time from the reported arrival times.
We applied HypoNet Nankai PS to these travel-time data. 
We did not apply the station corrections employed in the JMA analyses because those corrections were determined under the assumption of the JMA 1D velocity model.

The horizontal locations obtained by HypoNet Nankai PS show minor differences from the JMA catalog locations and moderate uncertainty, except for a few events located east of the observation network (Fig. \ref{fig:DONET_JMA} (a)(b)).
This indicates that the locations of most events are well constrained horizontally in both our results and the JMA catalog.
By contrast, hypocenter distributions along three cross sections show a marked contrast in depth patterns (Fig. \ref{fig:DONET_JMA} (c)(d)).

\begin{figure}[htbp]
\centering
\includegraphics[clip, width=\textwidth]{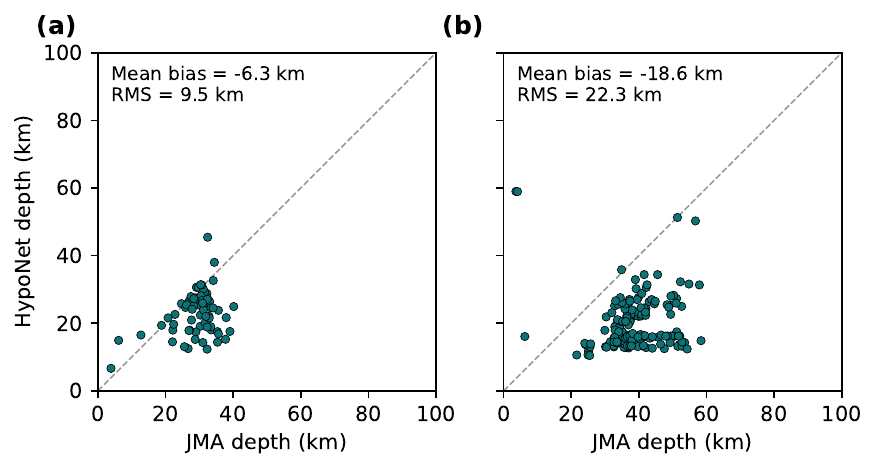}
\caption{Comparison of hypocenter depths estimated by HypoNet Nankai PS with those in the JMA catalog for the events shown in Fig. \ref{fig:DONET_JMA}.
The events are divided into two groups at a distance of -20\,km along the cross sections (dashed lines in Fig. \ref{fig:DONET_JMA} (a)(b)).
(a) Landward group (distances larger than -20\,km).
(b) Seaward group (distances smaller than -20\,km).
The diagonal line indicates equal depths.
The bias, defined as the mean of our depths minus the JMA depths, and the RMSD of the depth differences are shown for each group.}
\label{fig:hypo_depth_section0}
\end{figure}

We examine the depth differences in more detail, focusing on their spatial variation.
In the region landward of a distance of -20\,km in the cross sections, where the events lie beneath the DONET network and near the land, the estimated hypocenter depths are generally consistent with the JMA catalog depths. 
JMA hypocenters near the coast are generally expected to be more accurate than those farther offshore \citep{JMA_catalog_userguide}. 
This agreement suggests that, for events with favorable station geometry, the 3D velocity model used here captures travel-time variations that the JMA analysis effectively accommodates through a combination of JMA2020C, a 1D velocity model tuned for offshore stations in the Nankai Trough, and the station corrections.
In the region seaward of a distance of -20\,km, where the receivers are distant, the hypocenters estimated by HypoNet Nankai PS are substantially shallower than those in the JMA catalog. These events lie within the oceanic crust and upper mantle, as indicated by the plate boundary model proposed by \citet{Nakanishi2018}.
To clarify how the depth difference depends on the distance along the cross sections, we plotted our estimated depths against the JMA depths separately for two groups divided at a distance of -20\,km, namely the seaward group and the landward group (Fig. \ref{fig:hypo_depth_section0}). 
The dashed lines in Fig. \ref{fig:DONET_JMA} (a)(b) indicate this division. 
Both the bias and the RMS of the depth differences are considerably larger for the seaward group than for the landward group. 
For the seaward group, the bias, defined as the mean of our depths minus the JMA depths, is -18.6\,km. 
Although the analysis periods do not overlap, this bias is comparable to that reported by \citet{Yamamoto2022PEPS}, who relocated offshore earthquakes using DONET data and 3D velocity models and obtained hypocenters 10--15\,km shallower than the JMA catalog depths. 
At large offsets, travel times are expected to be dominated by long paths through complex three-dimensional structures, whose effects cannot be absorbed by station statics. 
We therefore interpret the offshore depth difference between our hypocenters and JMA's as an effect of velocity-model representation combined with station geometry and increased uncertainty. 
Our results show many hypocenters in the uppermost oceanic mantle. Although interpreting the seismicity distribution is beyond the scope of this study, \citet{Obana2005JGR} similarly located earthquakes in the uppermost oceanic mantle off the Kii Peninsula using temporary OBS observations and a 3D velocity structure, and associated this activity with compressive stress based on focal mechanisms.

Finally, we assessed the impact of incorporating S-wave travel times by repeating the analysis using only P-wave data (see Text S1 and Figure S2 for details). Omitting S-wave data left some events insufficiently constrained. Even for successfully located events, posterior uncertainties increased substantially: the 95th percentile of the horizontal posterior standard deviation increased from approximately 5\,km to 26\,km. These results demonstrate the practical benefit of incorporating S-wave travel-time modeling into HypoNet Nankai PS in stabilizing hypocenter estimates and reducing their uncertainties.



\section{Discussion and Conclusion}
\label{sec:conclusion}

In this study, we developed HypoNet Nankai PS, an extended hypocenter determination framework that enables rapid hypocenter determination using S-wave arrival times in addition to P-wave arrivals in a realistic 3D velocity structure of the Nankai Trough subduction zone. 
The principal methodological implication of this study is that extending PINN-based travel-time inference from P to S waves in a realistic subduction-zone structure is not simply a matter of training an additional network using the same configuration. 
The S-wave velocity model constructed here from empirical conversion relations contains very low velocities in unconsolidated near-seafloor sediments, producing substantially steeper travel-time gradients than those in the P-wave model. 
A training configuration developed for the P-wave field did not provide comparable accuracy for this more demanding S-wave field. 
By combining full-domain 3D sampling of source and collocation points with the Deeper-PINN architecture and the SOAP optimizer, the revised approach enabled a single global PINN for each wave type to reproduce FMM travel times with sufficient accuracy for hypocenter determination. 
Synthetic tests using FMM-generated arrival times showed that, where station geometry provided sufficient observational constraints, residual PINN travel-time approximation errors did not degrade the recovered hypocenter locations.
The larger errors and estimated uncertainties outside the station networks instead reflected the expected loss of constraint with increasing station distance and azimuthal gap, consistent with physical intuition.
The application to observed DONET data further demonstrated the practical viability of the framework. The resulting locations generally agreed with the JMA catalog beneath and near the DONET network, whereas the relocated depths farther offshore were substantially shallower and generally consistent with previous studies using DONET data and 3D velocity models. 
Because these analyses differed in the data sets, velocity models, and station-correction schemes, this comparison provides qualitative context rather than a direct assessment of relative accuracy or an isolated evaluation of the 3D velocity model. 
Finally, although the general importance of S-wave arrivals for hypocenter determination is well established, the P-only comparison quantified their practical contribution in the present DONET configuration, particularly in reducing their estimated uncertainties.

The compact storage requirement of the proposed framework is an important advantage of HypoNet Nankai PS. This point was already one of the strengths of the original HypoNet Nankai developed by A2026. However, that framework divided the model domain into one global region and 35 local subregions and trained separate neural networks for each region. 
In contrast, the improved training strategy introduced in this study enabled sufficiently accurate travel-time inference without such domain decomposition and the associated large number of neural networks. 
As a result, the data size of the P- and S-wave travel-time models of HypoNet Nankai PS is only 31 MB each and 63 MB in total.  
This is substantially smaller than the original HypoNet Nankai, for which the P-wave travel-time models alone required approximately 500 MB. This data size contrast is even more pronounced when compared with conventional travel-time tables. 
If FMM-based travel-time tables were constructed for the DONET and N-net observation networks, the estimated data volume would be approximately 80 GB for each of the P- and S-wave models. Thus, HypoNet Nankai PS provides a storage-efficient alternative to conventional precomputed travel-time tables.


HypoNet Nankai PS aims to enable a broad range of researchers to rapidly and conveniently determine hypocenters with results approaching the quality achieved through expert analyses accounting for three-dimensional velocity structures and/or appropriate station corrections. 
Conventional travel-time tables and station corrections must be recalculated and re-estimated, respectively, when new stations are added. 
In contrast, the PINN-based framework offers the flexibility to evaluate travel times for arbitrary source and receiver locations. This feature is particularly advantageous for seismic monitoring involving temporary campaign observations and also makes HypoNet Nankai PS well suited for preliminary numerical experiments designed to plan new observation campaigns. 
It should again be noted that the S-wave velocity model introduced in this study was derived through an empirical parameter conversion from the P-wave velocity structure and is not directly constrained by observed seismic data. 
Moreover, even when HypoNet Nankai PS is used, combining the framework with station corrections may be preferable in practical applications \citep{Toh2026EPS}. When an S-wave velocity model constrained by travel-time tomography using observed seismic data becomes available in the future, the PINN models should be updated accordingly to incorporate the improved structure.

\begin{table}[htbp]
\centering
\caption{Comparison of training algorithm settings between A2026 and HypoNet Nankai PS (this study).}
\label{tab:algorithm}
\begin{tabular}{lll}
\hline
Setting & A2026 & HypoNet Nankai PS \\
\hline
Collocation sampling & Full 3-D domain & Full 3-D domain \\
Source sampling & Earth's surface (2D manifold) & Full 3-D domain \\
Random-distance sampling & No & Yes \\
Optimizer & Adam & SOAP \\
NN architecture & Fully connected with RFF & Deeper-PINN with RFF\\
Domain decomposition & Yes & No \\
\hline
\end{tabular}
\end{table}

\section*{Declarations}

\section*{Availability of data and materials}
HypoNet Nankai PS will be available at \url{https://github.com/eagleray2020/HypoNet_Nankai_DeeperPINN_PS}. Access to the P-wave velocity structure can be requested through \url{https://www.jamstec.go.jp/obsmcs_db/j/Nankai_3D_model-j.html}. 
Similarly, the S-wave velocity structure will also be made available by JAMSTEC upon request, following acceptance of this manuscript.

\section*{Competing interests}
The authors declare that they have no competing interests.

\section*{Funding}
This study was supported by MEXT Project ``Earthquake Disaster Prevention Research Project for Damage Minimization and Rapid Response and Reconstruction from Great Earthquakes such as the Nankai Trough Earthquake'', JSPS KAKENHI Grant Number 24H01042, and JST SICORP Grant Number JPMJSC2401, Japan.

\section*{Authors' contributions}
RA designed the study, developed the software, performed the analyses, and drafted the manuscript. RA and GF aquired the fund. AN developed the S-wave velocity model. DB provided guidance on the design of the synthetic tests. SB and GF provided guidance on the design and interpretation of the real-data application. YN and GF provided scientific guidance throughout the project. All authors read and approved the final manuscript.

\section*{Authors' information}
Not applicable.

\acknowledgments{
The calculations were carried out using the Earth Simulator at JAMSTEC. AI-assisted tools were used for translation and grammar checking during the preparation of this manuscript.
}


\clearpage

\section*{Figure legends}

\begin{thebibliography}{54}
\providecommand{\natexlab}[1]{#1}
\providecommand{\url}[1]{\texttt{#1}}
\expandafter\ifx\csname urlstyle\endcsname\relax
  \providecommand{\doi}[1]{doi: #1}\else
  \providecommand{\doi}{doi: \begingroup \urlstyle{rm}\Url}\fi

\bibitem[Agata et~al.(2023)Agata, Shiraishi, and Fujie]{Agata2023}
Ryoichiro Agata, Kazuya Shiraishi, and Gou Fujie.
\newblock {Bayesian Seismic Tomography Based on Velocity-Space Stein
  Variational Gradient Descent for Physics-Informed Neural Network}.
\newblock \emph{IEEE Transactions on Geoscience and Remote Sensing},
  61:\penalty0 1--17, 2023.
\newblock \doi{10.1109/TGRS.2023.3295414}.

\bibitem[Agata et~al.(2026)Agata, Baba, Nakanishi, and Nakamura]{Agata2026SRL}
Ryoichiro Agata, Satoru Baba, Ayako Nakanishi, and Yasuyuki Nakamura.
\newblock {HypoNet Nankai: Rapid hypocenter determination tool for the Nankai
  Trough subduction zone using physics-informed neural networks}.
\newblock \emph{Seismological Research Letters}, 97\penalty0 (1):\penalty0
  524--536, 2026.

\bibitem[Aoi et~al.(2020)Aoi, Asano, Kunugi, Kimura, Uehira, Takahashi, Ueda,
  Shiomi, Matsumoto, and Fujiwara]{Aoi2020}
Shin Aoi, Youichi Asano, Takashi Kunugi, Takeshi Kimura, Kenji Uehira, Narumi
  Takahashi, Hideki Ueda, Katsuhiko Shiomi, Takumi Matsumoto, and Hiroyuki
  Fujiwara.
\newblock Mowlas: Nied observation network for earthquake, tsunami and volcano.
\newblock \emph{Earth, Planets and Space}, 72\penalty0 (1):\penalty0 1--31,
  2020.

\bibitem[Aoi et~al.(2023)Aoi, Takeda, Kunugi, Shinohara, Miyoshi, Uehira,
  Mochizuki, and Takahashi]{Aoi2023IEEE-UT}
Shin Aoi, Tetsuya Takeda, Takashi Kunugi, Masanao Shinohara, Takayuki Miyoshi,
  Kenji Uehira, Masashi Mochizuki, and Narumi Takahashi.
\newblock {Development and Construction of Nankai Trough Seafloor Observation
  Network for Earthquakes and Tsunamis: N-net}.
\newblock In \emph{2023 IEEE Underwater Technology (UT)}, pages 1--5. IEEE,
  2023.

\bibitem[Arnulf et~al.(2022)Arnulf, Bassett, Harding, Kodaira, Nakanishi, and
  Moore]{Arnulf2022NatGeo}
Adrien~F Arnulf, Dan Bassett, Alistair~J Harding, Shuichi Kodaira, Ayako
  Nakanishi, and Gregory Moore.
\newblock Upper-plate controls on subduction zone geometry, hydration and
  earthquake behaviour.
\newblock \emph{Nature Geoscience}, 15\penalty0 (2):\penalty0 143--148, 2022.

\bibitem[Bassett et~al.(2022)Bassett, Arnulf, Kodaira, Nakanishi, Harding, and
  Moore]{Bassett2022JGR-SE}
Dan Bassett, Adrien Arnulf, Shuichi Kodaira, Ayako Nakanishi, Alistair Harding,
  and Gregory Moore.
\newblock {Crustal structure of the Nankai subduction zone revealed by two
  decades of onshore-offshore and ocean-bottom seismic data: Implications for
  the dimensions and slip behavior of the seismogenic zone}.
\newblock \emph{Journal of Geophysical Research: Solid Earth}, 127\penalty0
  (10):\penalty0 e2022JB024992, 2022.

\bibitem[Bassett et~al.(2025)Bassett, Henrys, Tozer, van Avendonk, Gase, Bangs,
  Kodaira, Okaya, Jacobs, Sutherland, Seebeck, Barker, Fujie, Arai, Seaward,
  Mochizuki, Savage, Stern, and Luckie]{Bassett2025JGR-SE}
Dan Bassett, Stuart Henrys, Brook Tozer, Harm van Avendonk, Andrew Gase, Nathan
  Bangs, Shuichi Kodaira, David Okaya, Katie Jacobs, Rupert Sutherland, Hannu
  Seebeck, Dan Barker, Gou Fujie, Ryuta Arai, Anya Seaward, Kimi Mochizuki,
  Martha Savage, Tim Stern, and Thomas Luckie.
\newblock Crustal structure of {the Hikurangi} subduction zone revealed by four
  decades of onshore-offshore seismic data: Implications for the dimensions and
  slip behavior of the seismogenic zone.
\newblock \emph{Journal of Geophysical Research: Solid Earth}, 130\penalty0
  (1):\penalty0 e2024JB030268, 2025.
\newblock \doi{https://doi.org/10.1029/2024JB030268}.
\newblock URL
  \url{https://agupubs.onlinelibrary.wiley.com/doi/abs/10.1029/2024JB030268}.

\bibitem[Brocher(2005)]{Brocher2005}
Thomas~M Brocher.
\newblock {Empirical relations between elastic wavespeeds and density in the
  Earth's crust}.
\newblock \emph{Bulletin of the seismological Society of America}, 95\penalty0
  (6):\penalty0 2081--2092, 2005.

\bibitem[Ganster(2023)]{Ganster2023eikonalfm}
Kevin Ganster.
\newblock {eikonalfm: Python module for solving the Eikonal equation using the
  fast marching method}.
\newblock \url{https://github.com/kevinganster/eikonalfm}, 2023.
\newblock Accessed: 2025-08-14.

\bibitem[{GEBCO Bathymetric Compilation Group 2023}(2023)]{Gebco2023}
{GEBCO Bathymetric Compilation Group 2023}.
\newblock {GEBCO 2023 Grid - a continuous terrain model of the global oceans
  and land}, 2023.

\bibitem[Giroux(2021)]{Giroux2021}
Bernard Giroux.
\newblock ttcrpy: A python package for traveltime computation and raytracing.
\newblock \emph{SoftwareX}, 16:\penalty0 100834, 2021.

\bibitem[Grubas et~al.(2023)Grubas, Duchkov, and Loginov]{Grubas2023}
Serafim Grubas, Anton Duchkov, and Georgy Loginov.
\newblock {Neural Eikonal solver: Improving accuracy of physics-informed neural
  networks for solving eikonal equation in case of caustics}.
\newblock \emph{Journal of Computational Physics}, 474:\penalty0 111789, 2023.

\bibitem[Hamilton(1979)]{Hamilton1979JASA}
Edwin~L Hamilton.
\newblock {$V_p/V_s$ and Poisson's ratios in marine sediments and rocks}.
\newblock \emph{The Journal of the Acoustical Society of America}, 66\penalty0
  (4):\penalty0 1093--1101, 1979.

\bibitem[Hennigh et~al.(2021)Hennigh, Narasimhan, Nabian, Subramaniam,
  Tangsali, Fang, Rietmann, Byeon, and Choudhry]{Hennigh2021}
Oliver Hennigh, Susheela Narasimhan, Mohammad~Amin Nabian, Akshay Subramaniam,
  Kaustubh Tangsali, Zhiwei Fang, Max Rietmann, Wonmin Byeon, and Sanjay
  Choudhry.
\newblock Nvidia simnet™: An ai-accelerated multi-physics simulation
  framework.
\newblock In \emph{International conference on computational science}, pages
  447--461. Springer, 2021.

\bibitem[Hirata and Matsu'ura(1987)]{Hirata1987}
Naoshi Hirata and Mitsuhiro Matsu'ura.
\newblock Maximum-likelihood estimation of hypocenter with origin time
  eliminated using nonlinear inversion technique.
\newblock \emph{Physics of the Earth and Planetary Interiors}, 47:\penalty0
  50--61, 1987.

\bibitem[{JAMSTEC}(2004)]{JAMSTEC2004}
{JAMSTEC}.
\newblock {JAMSTEC Seismic Survey Database}, 2004.
\newblock \url{https://doi.org/10.17596/0002069}, Accessed 10 June, 2024.

\bibitem[{Japan Meteorological Agency}(2026)]{JMA_catalog_userguide}
{Japan Meteorological Agency}.
\newblock User's guide: Monthly report on earthquakes and volcanoes in japan
  (catalog).
\newblock
  \url{https://www.data.jma.go.jp/eqev/data/bulletin/catalog/notes_e.html},
  2026.
\newblock Accessed 17 June 2026.

\bibitem[Jiang et~al.(2025)Jiang, Hou, Ye, and Xia]{Jiang2025ASC}
Feilong Jiang, Xiaonan Hou, Jianqiao Ye, and Min Xia.
\newblock {Deeper-PINNs: Unlocking the power of deep physics-informed neural
  networks}.
\newblock \emph{Applied Soft Computing}, page 114048, 2025.

\bibitem[Johnson and Vincent(2002)]{Johnson2002BSSA}
Michelle Johnson and Carolynn Vincent.
\newblock {Development and testing of a 3D velocity model for improved event
  location: a case study for the India-Pakistan region}.
\newblock \emph{Bulletin of the Seismological Society of America}, 92\penalty0
  (8):\penalty0 2893--2910, 2002.

\bibitem[Julian et~al.(1977)Julian, Gubbins, et~al.]{Julian1977JG}
BR~Julian, Db~Gubbins, et~al.
\newblock Three-dimensional seismic ray tracing.
\newblock \emph{Journal of Geophysics}, 43\penalty0 (1):\penalty0 95--113,
  1977.

\bibitem[Kaneda et~al.(2015)Kaneda, Kawaguchi, Araki, Matsumoto, Nakamura,
  Kamiya, Ariyoshi, Hori, Baba, and Takahashi]{Kaneda2015}
Y~Kaneda, K~Kawaguchi, E~Araki, H~Matsumoto, T~Nakamura, S~Kamiya, K~Ariyoshi,
  T~Hori, T~Baba, and N~Takahashi.
\newblock {Development and application of an advanced ocean floor network
  system for megathrust earthquakes and tsunamis}.
\newblock In \emph{Seafloor observatories}, pages 643--662. Springer, 2015.

\bibitem[Katsumata(2015)]{Katsumata2015BSSA}
Akio Katsumata.
\newblock {Fast hypocenter determination in an inhomogeneous velocity structure
  using a 3D travel-time table}.
\newblock \emph{Bulletin of the Seismological Society of America}, 105\penalty0
  (6):\penalty0 3203--3208, 2015.

\bibitem[Kingma and Ba(2015)]{Kingma2015}
Diederik~P. Kingma and Jimmy Ba.
\newblock {Adam: A method for stochastic optimization}.
\newblock In \emph{International Conference on Learning Representations}, 2015.

\bibitem[Liu and Nocedal(1989)]{Liu1989}
Dong~C Liu and Jorge Nocedal.
\newblock {On the limited memory BFGS method for large scale optimization}.
\newblock \emph{Mathematical programming}, 45\penalty0 (1):\penalty0 503--528,
  1989.

\bibitem[Liu et~al.(2025)Liu, Nakanishi, Fujie, Tonegawa, Nakamura, Obana, and
  Kodaira]{Liu2025PNAS}
Xin Liu, Ayako Nakanishi, Gou Fujie, Takashi Tonegawa, Yasuyuki Nakamura,
  Koichiro Obana, and Shuichi Kodaira.
\newblock {Abundant fluids in southern Kumano Basin linked to fluid source and
  slow earthquakes at the plate boundary in the Nankai Trough}.
\newblock \emph{Proceedings of the National Academy of Sciences}, 122\penalty0
  (32):\penalty0 e2423440122, 2025.

\bibitem[Lomax et~al.(2000)Lomax, Virieux, Volant, and
  Berge-Thierry]{Lomax2000}
Anthony Lomax, Jean Virieux, Philippe Volant, and Catherine Berge-Thierry.
\newblock Probabilistic earthquake location in 3d and layered models:
  Introduction of a metropolis-gibbs method and comparison with linear
  locations.
\newblock \emph{Advances in seismic event location}, pages 101--134, 2000.

\bibitem[Matsumura et~al.(2006)Matsumura, Ito, Kimura, Obara, Sekiguchi, Hori,
  and Kasahara]{Matsumura2006Zisin}
Minoru Matsumura, Yoshihiro Ito, Hisanori Kimura, Kazushige Obara, Shoji
  Sekiguchi, Sadaki Hori, and Keiji Kasahara.
\newblock {Development of Accurate and Quick Analysis System for Source
  Parameters (AQUA) (in Japanese with English abstract)}.
\newblock \emph{Zisin (Journal of the Seismological Society of Japan. 2nd
  ser.)}, 59\penalty0 (2):\penalty0 167--184, 2006.
\newblock \doi{10.4294/zisin.59.167}.

\bibitem[{MEXT and JAMSTEC}(2022)]{MEXT2022NankaiBosai}
{MEXT and JAMSTEC}.
\newblock {FY2021 Report: Research Project for Disaster Prevention on the Great
  Earthquakes along the Nankai Trough}.
\newblock
  \url{https://www.jishin.go.jp/database/project_report/nankai-bosai/nankai-bosai-r03/},
  2022.
\newblock Accessed: June 26, 2026, in Japanese.

\bibitem[Moser(1991)]{Moser1991Geophysics}
TJ~Moser.
\newblock Shortest path calculation of seismic rays.
\newblock \emph{Geophysics}, 56\penalty0 (1):\penalty0 59--67, 1991.

\bibitem[Nakanishi et~al.(2018)Nakanishi, Takahashi, Yamamoto, Takahashi,
  Citak, Nakamura, Obana, Kodaira, and Kaneda]{Nakanishi2018}
Ayako Nakanishi, Narumi Takahashi, Yojiro Yamamoto, Tsutomu Takahashi,
  Seckin~Ozgur Citak, Takeshi Nakamura, Koichiro Obana, Shuichi Kodaira, and
  Yoshiyuki Kaneda.
\newblock {Three-dimensional plate geometry and P-wave velocity models of the
  subduction zone in SW Japan: Implications for seismogenesis}.
\newblock \emph{Geology and Tectonics of Subduction Zones: A Tribute to Gaku
  Kimura}, 534:\penalty0 69, 2018.

\bibitem[Nakano et~al.(2015)Nakano, Nakamura, and Kaneda]{Nakano2015}
Masaru Nakano, Takeshi Nakamura, and Yoshiyuki Kaneda.
\newblock {Hypocenters in the Nankai Trough Determined by Using Data from Both
  Ocean-Bottom and Land Seismic Networks and a 3D Velocity Structure Model:
  Implications for Seismotectonic Activity}.
\newblock \emph{Bulletin of the Seismological Society of America}, 105\penalty0
  (3):\penalty0 1594--1605, 2015.

\bibitem[Naoi et~al.(2024)Naoi, Tamaribuchi, Shimojo, Katoh, and
  Ohyanagi]{Naoi2024EPS}
Makoto Naoi, Koji Tamaribuchi, Kengo Shimojo, Shinya Katoh, and Shukei
  Ohyanagi.
\newblock {Neural phase picker trained on the Japan meteorological agency
  unified earthquake catalog}.
\newblock \emph{Earth, Planets and Space}, 76\penalty0 (1):\penalty0 150, 2024.

\bibitem[Obana et~al.(2005)Obana, Kodaira, and Kaneda]{Obana2005JGR}
Koichiro Obana, Shuichi Kodaira, and Yoshiyuki Kaneda.
\newblock {Seismicity in the incoming/subducting Philippine Sea plate off the
  Kii Peninsula, central Nankai trough}.
\newblock \emph{Journal of Geophysical Research: Solid Earth}, 110\penalty0
  (B11), 2005.

\bibitem[Pavlis et~al.(2012)Pavlis, Holmes, Kenyon, and Factor]{Pavlis2012}
Nikolaos~K Pavlis, Simon~A Holmes, Steve~C Kenyon, and John~K Factor.
\newblock {The development and evaluation of the Earth Gravitational Model 2008
  (EGM2008)}.
\newblock \emph{Journal of geophysical research: solid earth}, 117\penalty0
  (B4), 2012.

\bibitem[Raissi et~al.(2019)Raissi, Perdikaris, and Karniadakis]{Raissi2019}
Maziar Raissi, Paris Perdikaris, and George~E Karniadakis.
\newblock {Physics-informed neural networks: A deep learning framework for
  solving forward and inverse problems involving nonlinear partial differential
  equations}.
\newblock \emph{Journal of Computational physics}, 378:\penalty0 686--707,
  2019.

\bibitem[Ross et~al.(2026)Ross, Wilding, Azizzadenesheli, and
  Kato]{Ross2026JGR-SE}
Zachary~E Ross, John~D Wilding, Kamyar Azizzadenesheli, and Aitaro Kato.
\newblock {SPIDER: Scalable probabilistic inference for differential earthquake
  relocation}.
\newblock \emph{Journal of Geophysical Research: Solid Earth}, 131\penalty0
  (3):\penalty0 e2025JB032769, 2026.

\bibitem[Sethian(1996)]{Sethian1996}
James~A Sethian.
\newblock A fast marching level set method for monotonically advancing fronts.
\newblock \emph{proceedings of the National Academy of Sciences}, 93\penalty0
  (4):\penalty0 1591--1595, 1996.

\bibitem[Shaw et~al.(2015)Shaw, Plesch, Tape, Suess, Jordan, Ely, Hauksson,
  Tromp, Tanimoto, Graves, et~al.]{Shaw2015EPSL}
John~H Shaw, Andreas Plesch, Carl Tape, M~Peter Suess, Thomas~H Jordan,
  Geoffrey Ely, Egill Hauksson, Jeroen Tromp, Toshiro Tanimoto, Robert Graves,
  et~al.
\newblock Unified structural representation of the southern california crust
  and upper mantle.
\newblock \emph{Earth and Planetary Science Letters}, 415:\penalty0 1--15,
  2015.

\bibitem[Smith et~al.(2021)Smith, Azizzadenesheli, and Ross]{Smith2021}
Jonathan~D. Smith, Kamyar Azizzadenesheli, and Zachary~E. Ross.
\newblock Eikonet: Solving the eikonal equation with deep neural networks.
\newblock \emph{IEEE Transactions on Geoscience and Remote Sensing},
  59\penalty0 (12):\penalty0 10685--10696, 2021.
\newblock \doi{10.1109/TGRS.2020.3039165}.

\bibitem[Smith et~al.(2022)Smith, Ross, Azizzadenesheli, and Muir]{Smith2022}
Jonthan~D Smith, Zachary~E Ross, Kamyar Azizzadenesheli, and Jack~B Muir.
\newblock {HypoSVI: Hypocentre inversion with Stein variational inference and
  physics informed neural networks}.
\newblock \emph{Geophysical Journal International}, 228\penalty0 (1):\penalty0
  698--710, 2022.

\bibitem[Takahashi et~al.(2002)Takahashi, Kodaira, Nakanishi, Park, Miura,
  Tsuru, Kaneda, Suyehiro, Kinoshita, Hirata, et~al.]{Takahashi2002JGR-SE}
Narumi Takahashi, Shuichi Kodaira, Ayako Nakanishi, Jin-Oh Park, Seiichi Miura,
  Tetsuro Tsuru, Yoshiyuki Kaneda, Kiyoshi Suyehiro, Hajimu Kinoshita, Naoshi
  Hirata, et~al.
\newblock {Seismic structure of western end of the Nankai trough seismogenic
  zone}.
\newblock \emph{Journal of Geophysical Research: Solid Earth}, 107\penalty0
  (B10):\penalty0 ESE--2, 2002.

\bibitem[Tancik et~al.(2020)Tancik, Srinivasan, Mildenhall, Fridovich-Keil,
  Raghavan, Singhal, Ramamoorthi, Barron, and Ng]{Tancik2020}
Matthew Tancik, Pratul Srinivasan, Ben Mildenhall, Sara Fridovich-Keil, Nithin
  Raghavan, Utkarsh Singhal, Ravi Ramamoorthi, Jonathan Barron, and Ren Ng.
\newblock {Fourier features let networks learn high frequency functions in low
  dimensional domains}.
\newblock \emph{Advances in Neural Information Processing Systems},
  33:\penalty0 7537--7547, 2020.

\bibitem[Taufik et~al.(2023)Taufik, Waheed, and Alkhalifah]{Taufik2023SciRep}
Mohammad~H Taufik, Umair~bin Waheed, and Tariq~A Alkhalifah.
\newblock {A neural network based global traveltime function (GlobeNN)}.
\newblock \emph{Scientific Reports}, 13\penalty0 (1):\penalty0 7179, 2023.

\bibitem[Toh et~al.(2026)Toh, Baba, Nakanishi, Agata, Nakano, Yamamoto, and
  Araki]{Toh2026EPS}
Akiko Toh, Satoru Baba, Ayako Nakanishi, Ryoichiro Agata, Masaru Nakano, Yojiro
  Yamamoto, and Eiichiro Araki.
\newblock {Hypocenter Location Using Submarine DAS in the Nankai Trough
  Requires Correct Phase Identification, Station Corrections, and Reliable 3D
  Structures}.
\newblock \emph{Earth, Planets and Space}, 2026.
\newblock under review.

\bibitem[Tonegawa et~al.(2017)Tonegawa, Araki, Kimura, Nakamura, Nakano, and
  Suzuki]{Tonegawa2017NatComms}
Takashi Tonegawa, Eiichiro Araki, Toshinori Kimura, Takeshi Nakamura, Masaru
  Nakano, and Kensuke Suzuki.
\newblock Sporadic low-velocity volumes spatially correlate with shallow very
  low frequency earthquake clusters.
\newblock \emph{Nature Communications}, 8\penalty0 (1):\penalty0 2048, 2017.

\bibitem[Um and Thurber(1987)]{Um1987BSSA}
Junho Um and Clifford Thurber.
\newblock A fast algorithm for two-point seismic ray tracing.
\newblock \emph{Bulletin of the Seismological Society of America}, 77\penalty0
  (3):\penalty0 972--986, 1987.

\bibitem[Vyas et~al.(2025)Vyas, Morwani, Zhao, Shapira, Brandfonbrener, Janson,
  and Kakade]{Vyas2025ICLR}
Nikhil Vyas, Depen Morwani, Rosie Zhao, Itai Shapira, David Brandfonbrener,
  Lucas Janson, and Sham~M Kakade.
\newblock {SOAP: Improving and Stabilizing Shampoo using Adam for Language
  Modeling}.
\newblock In \emph{The Thirteenth International Conference on Learning
  Representations}, 2025.

\bibitem[Waheed et~al.(2021)Waheed, Haghighat, Alkhalifah, Song, and
  Hao]{Waheed2021PINNeik}
Umair~Bin Waheed, Ehsan Haghighat, Tariq Alkhalifah, Chao Song, and Qi~Hao.
\newblock {PINNeik: Eikonal solution using physics-informed neural networks}.
\newblock \emph{Computers \& Geosciences}, 155:\penalty0 104833, 2021.

\bibitem[Wang et~al.(2025)Wang, bhartari, Li, and Perdikaris]{Wang2025NeurIPS}
Sifan Wang, Ananyae~Kumar bhartari, Bowen Li, and Paris Perdikaris.
\newblock {Gradient Alignment in Physics-informed Neural Networks: A
  Second-Order Optimization Perspective}.
\newblock In \emph{The Thirty-ninth Annual Conference on Neural Information
  Processing Systems}, 2025.
\newblock URL \url{https://openreview.net/forum?id=iweeVl1RHU}.

\bibitem[White et~al.(2020)White, Fang, Nakata, and Ben-Zion]{White2020SRL}
Malcolm~CA White, Hongjian Fang, Nori Nakata, and Yehuda Ben-Zion.
\newblock {PyKonal: a Python package for solving the eikonal equation in
  spherical and Cartesian coordinates using the fast marching method}.
\newblock \emph{Seismological Research Letters}, 91\penalty0 (4):\penalty0
  2378--2389, 2020.

\bibitem[Yamamoto et~al.(2013)Yamamoto, Obana, Takahashi, Nakanishi, Kodaira,
  and Kaneda]{Yamamoto2013}
Yojiro Yamamoto, Koichiro Obana, Tsutomu Takahashi, Ayako Nakanishi, Shuichi
  Kodaira, and Yoshiyuki Kaneda.
\newblock {Imaging of the subducted Kyushu-Palau Ridge in the Hyuga-nada
  region, western Nankai Trough subduction zone}.
\newblock \emph{Tectonophysics}, 589:\penalty0 90--102, 2013.

\bibitem[Yamamoto et~al.(2014)Yamamoto, Obana, Takahashi, Nakanishi, Kodaira,
  and Kaneda]{Yamamoto2014}
Yojiro Yamamoto, Koichiro Obana, Tsutomu Takahashi, Ayako Nakanishi, Shuichi
  Kodaira, and Yoshiyuki Kaneda.
\newblock {Seismicity and structural heterogeneities around the western Nankai
  Trough subduction zone, southwestern Japan}.
\newblock \emph{Earth and Planetary Science Letters}, 396:\penalty0 34--45,
  2014.

\bibitem[Yamamoto et~al.(2022)Yamamoto, Yada, Ariyoshi, Hori, and
  Takahashi]{Yamamoto2022PEPS}
Yojiro Yamamoto, Shuichiro Yada, Keisuke Ariyoshi, Takane Hori, and Narumi
  Takahashi.
\newblock {Seismicity distribution in the Tonankai and Nankai seismogenic zones
  and its spatiotemporal relationship with interplate coupling and slow
  earthquakes}.
\newblock \emph{Progress in Earth and Planetary Science}, 9\penalty0
  (1):\penalty0 1--20, 2022.

\bibitem[Zhang and Thurber(2003)]{Zhang2003}
Haijiang Zhang and Clifford~H Thurber.
\newblock {Double-difference tomography: The method and its application to the
  Hayward fault, California}.
\newblock \emph{Bulletin of the Seismological Society of America}, 93\penalty0
  (5):\penalty0 1875--1889, 2003.

\end{thebibliography}
\end{document}